\documentclass[journal,10pt,twocolumn]{IEEEtran}

\usepackage[cmex10]{amsmath}
\usepackage{amssymb}
\usepackage{amsthm}
\usepackage{graphicx}
\usepackage{color}
\usepackage{bm}
\usepackage{bbm}
\usepackage{url}
\usepackage{array}
\usepackage{multicol}
\usepackage{algorithm}
\usepackage[colorlinks]{hyperref}
\usepackage{url}
\usepackage{algpseudocode}

\theoremstyle{remark}

\usepackage{array}
\makeatletter
\newcommand{\thickhline}{%
    \noalign {\ifnum 0=`}\fi \hrule height 1pt
    \futurelet \reserved@a \@xhline
}
\newcolumntype{"}{@{\hskip\tabcolsep\vrule width 1pt\hskip\tabcolsep}}
\makeatother

\begin{document}

\title{Turbocharging Fluid Antenna Multiple Access}

\author{\IEEEauthorblockN{Noor Waqar, 
		                           Kai-Kit Wong, \emph{Fellow, IEEE},
		                           Chan-Byoung Chae, \emph{Fellow, IEEE}, and 
		                           Ross Murch, \emph{Fellow, IEEE}
}

\vspace{-8mm}


\thanks{The work of K. K. Wong is supported by the Engineering and Physical Sciences Research Council (EPSRC) under grant EP/W026813/1.}
\thanks{The work of R. Murch was supported by the Hong Kong Research Grants Council Area of Excellence Grant AoE/E-601/22-R.}
\thanks{The work of C. B. Chae is supported by the Institute for Information and Communication Technology Planning and Evaluation (IITP)/NRF grant funded by the Ministry of Science and ICT (MSIT), South Korea, under Grant RS-2024-00428780 and 2022R1A5A1027646.}

\thanks{N. Waqar and K. K. Wong are with the Department of Electronic and Electrical Engineering, University College London, London WC1E 7JE, United Kingdom. K. K. Wong is also affiliated with Yonsei Frontier Laboratory, Yonsei University, Seoul, 03722, Korea.}
\thanks{C. B. Chae is with School of Integrated Technology, Yonsei University, Seoul, 03722, Korea.}
\thanks{R. Murch is with the Department of Electronic and Computer Engineering and Institute for Advanced Study (IAS), Hong Kong University of Science and Technology, Clear Water Bay, Hong Kong SAR, China.}

\thanks{ Corresponding author: Kai-Kit Wong (e-mail: $\rm kai\text{-}kit.wong@ucl.ac.uk$).}
}
\maketitle

\begin{abstract}
Based on our current understanding, extreme massive access over the same physical channel is only possible if an extra-large multiple-input multiple-output (XL-MIMO) antenna is used at the base station (BS) and instantaneous channel state information (CSI) is known at the BS side for precoding design. This casts doubt on scalability and challenges in device-to-device situations in which there is not a centralized, optimized BS for transmitting the user signals. To address this problem, we revisit the massive connectivity challenge by considering the case where no CSI is available at the BS and no precoding is used. In this situation, inter-user interference (IUI) mitigation can only be performed at the user terminal (UT) side. Leveraging the position flexibility of fluid antenna system (FAS), we adopt a fluid antenna multiple access (FAMA) approach that exploits the interference signal fluctuation in the spatial domain. Specifically, we assume that we have $N$ spatially correlated received signals per symbol duration from FAS. Our main approach uses a simple heuristic port shortlisting method that identifies promising ports to obtain favourable received signals that can be combined via maximum ratio combining (MRC) to form the received output signal for final detection. On top of this, a pre-trained deep joint source-channel coding (JSCC) scheme is employed, which  together with a diffusion-based denoising model (MixDDPM) at the UT side, can improve the IUI immunity. We refer to the proposed scheme as {\em turbo} FAMA. Simulation results show that with a physical FAS size of $20$ wavelengths at each UT transmitting quaternary phase shift keying (QPSK) symbols, fast FAMA can support $50$ users while turbo FAMA can handle up to $200$ users if the required symbol error rate (SER) is $10^{-2}$. If a higher error tolerance is acceptable, say SER at $0.1$, turbo FAMA can even serve up to $1000$ users but fast FAMA is only able to handle $160$ users, all remarkably achieved without CSI at the BS.
\end{abstract}


\vspace{-2mm}
\section{Introduction}
\subsection{Background}
\IEEEPARstart{S}{pectrum} is precious and the success of wireless communications technologies depends highly upon how much we can reuse our frequency bands \cite{Andrews-2011}. In traditional orthogonal multiple access (OMA) approaches such as frequency-division multiple access (FDMA), there is an obvious trade-off between the number of supportable users and the bandwidth each user is allowed to have. Frequency reuse or more generally multiple access on the same physical channel, on the other hand, hints at an exciting possibility of having both at the same time. Indeed, the highly anticipated sixth generation (6G) has the mission to deliver data rates of up to $100~{\rm Gbps}$, with peak rates reaching $1~{\rm Tbps}$, a connection density of $10^{7}$ per km$^{2}$, latency under $1~{\rm ms}$, and many other ambitious targets \cite{Zhang-2019,Saad-2020,Tariq-2020}. It is worth pointing out that latency in wireless networks is primarily due to the complexity of the protocol stack, and not the propagation delay. If multiple access techniques are highly complex, then latency will likely be high. Currently, there is a multiple access techniques (MAT) Industry Specification Group (ISG) set up in the European Telecommunications Standards Institute (ETSI) to explore novel multiuser access strategies \cite{ETSI_RSMA}.

In today's technologies, multiuser multiple-input multiple-output (MIMO) has been the most celebrated multiple access technique. The first multiuser MIMO work dates back in 2000 by Wong {\em et al.}~\cite{Wong-2000} and since then, numerous results appeared, e.g., \cite{Wong-2002,Wong-2003,Vishwanath-2003,Choi-2004,Spencer-2004}. Multiuser MIMO permits multiple users to share the same physical channel by focusing the signal beams onto the desired users and placing nulls at unintended receivers---a technique called {\em precoding}. With decent channel state information (CSI) at the base station (BS) side, multiuser MIMO promises to handle as many users in the downlink as the number of BS antennas. In recent generations, multiuser MIMO has evolved into massive MIMO, promoting the use of an excessive number of BS antennas (typically $6$ times the number of users) to simplify precoding design \cite{Marzetta-2010,Larsson-2014}. The idea is elegant but in practice, massive MIMO reverts back to multiuser minimum mean-square-error (MMSE) precoding in the fifth generation (5G) \cite{Villalonga-2022} and the extra antennas are there to provide more degree-of-freedom (DoF) to avoid inter-user interference (IUI) more effectively. To realize 6G, a natural step will be to deploy even more BS antennas in the form of extra-large MIMO (XL-MIMO) \cite{Wang-2024mimo} or a cell-free setup \cite{Ngo-2017}. However, the overhead for CSI acquisition and the huge power consumption of a massive array call for a different design.

Besides MIMO, non-orthogonal multiple access (NOMA) and rate-splitting multiple access (RSMA) both have spurred enormous interest in recent years \cite{liu2024road}. In NOMA, user signals are superimposed and IUI is subtracted at the user side using successive interference cancellation (SIC), see e.g., \cite{Saito-2013,Ping-2006,Dai-2015,Liu-2022}. NOMA does not scale nicely with the number of users since SIC becomes prohibitively complex very quickly. By contrast, RSMA manages the IUI by splitting the messages into common and private parts, and each user first decodes the common message by treating other signals as noise and then decodes its private message after subtracting the interference from the common message \cite{Mao-2022,Clerckx-2023}. Although SIC is still needed in RSMA, each user only needs to cancel the interference from the common message. As such, RSMA is believed to be more scalable than NOMA. Nevertheless, both schemes require CSI at the BS to perform power control as well as user clustering for the best performance. Due to high complexity, the majority of results are limited to only a few users.

\vspace{2mm}
\begin{quote}
\begin{center}
{\em ``Can space-division multiple-access be much simpler, at least conceptually?''}
\end{center}
\end{quote}
\vspace{2mm}

This was the question asked in \cite{wong2022FAMA} when pursuing a new multiple access method that is hopefully more straightforward, a lot more scalable, and does not require complex interference management, unlike multiuser MIMO, NOMA and RSMA. In this paper, we aim to shed light on the answer to a similar, but more specific question as follows:

\vspace{2mm}
\begin{quote}
\begin{center}
{\em ``Can extreme massive multiple access be accomplished without CSI at the transmitter side, without centralized interference management and without SIC at the user side?''}
\end{center}
\end{quote}
\vspace{2mm}

Based on the results in \cite{wong2022FAMA}, the answer is quite positive. Specifically, \cite{wong2022FAMA} proposed the concept of fluid antenna multiple access (FAMA). Through antenna position reconfiguration, a user can access the position, a.k.a.~port, where the ratio of the desired signal energy to the energy of the sum-interference plus noise signal is maximized on a per-symbol basis, dealing with IUI entirely from the receiver side without interference management. This idea was later called fast FAMA \cite{Wong-2024ffama}.

The origin of the idea came from an earlier concept, referred to as fluid antenna system (FAS) \cite{Wong-2022fcn,Wong-PartI2023,New-2024tut}. FAS was first envisaged by Wong {\em et al.}~in 2020 \cite{Wong-2020fas,Wong-2021fas} and FAS broadly represents the new generation of reconfigurable antennas that uses shape and position flexibility in antennas to empower the physical layer. Most recently, \cite{Lu-2025} and \cite{zhang2024pixel} explained the concept of FAS from the viewpoint of electromagnetics. A lot of opportunities for using artificial intelligence (AI) techniques in FAS were discussed in \cite{Wang-aifas2024}. It is worth emphasizing that FAS is not restricted to any specific implementation methods. It can be realized by a variety of ways, such as liquid-based antennas \cite{Huang-2021,shen2024design,Shamim-2025}, movable array \cite{basbug2017design}, metamaterial-based antennas \cite{Liu-2025arxiv}, and reconfigurable pixels \cite{zhang2024pixel,Shen-2024}.

Following the seminal work in \cite{Wong-2020fas,Wong-2021fas}, FAS has been extensively investigated in point-to-point communication scenarios. For instance, \cite{Khammassi2023} proposed an eigenvalue-based channel model to facilitate diversity analysis for FAS channels. Then the authors in \cite{New2023fluid} presented a method to compute the diversity order based on the model in \cite{Khammassi2023}. A spatial block-correlation model that maintains analytical tractability and approximates accurately the model for FAS channels was proposed in \cite{Espinosa-2024}. Subsequent research further extended the analysis to Nakagami fading channels \cite{Vega2023asimple,Vega2023novel}, $\alpha$-$\mu$ fading channels \cite{Alvim2023on} and even arbitrarily distributed fading channels \cite{Ghadi-2023}. Copula theory has also been shown to be a powerful tool for the analysis of FAS channels  \cite{Ghadi-2024}. Recent efforts have also found FAS applied in MIMO systems and \cite{new2023information} uncovered the diversity-multiplexing trade-off of such systems. Finding the best antenna positions in FAS evidently requires CSI to be known. Efforts for estimating the CSI of FAS can be found in \cite{xu2024channel,zhang2023successive,Xu-2025ce,10751774}.

A wide range of applications have already been illustrated to benefit from the deployment of FAS. For instance, FAS can introduce signal randomness to enhance physical layer security \cite{Tang-2023,Xu-2024pls,Ghadi-2024dec}. In \cite{Skouroumounis-2023}, FAS was used to synergize full-duplex communications. Reconfigurable intelligent surface (RIS) can also combine with FAS for promising performance \cite{10539238,salem2025first}. Moreover, in \cite{Yang-2024pim,Zhu-2025ris}, it was shown that FAS allows a new form of index modulation to trade-off diversity performance and communication rate. Recent trends also see efforts to use FAS for expanding the communication-sensing achievability region for integrated sensing and communication (ISAC) \cite{Wang-2024isac,zhou2024fasisac,Zou-2024,Meng-2025}. It is worth stating that FAS broadly includes the emerging topics of movable antennas \cite{zhu2024historical}, flexible-position MIMO \cite{10480333}, and pinching antennas \cite{Yang-2025pa} as well.

\vspace{-2mm}
\subsection{Literature Review of FAMA}
Going back to multiple access, FAMA indeed offers a very different understanding of how IUI can be mitigated---the one in which centralized interference management is not necessary and multipath fading becomes a desirable factor. In spite of a very positive outlook in \cite{wong2022FAMA}, the authors did little to provide any clarity of how the best port can be found to maximize the ratio of the desired signal energy to the energy of the sum-interference plus noise signal on a per-symbol basis. Later in \cite{Wong-2024ffama}, an attempt was made to overcome this problem, but not without considerable performance loss. Furthermore, antenna positioning switching as fast as symbol rate, albeit doable via designs in \cite{Liu-2025arxiv,zhang2024pixel}, is by no means trivial, which has led to the slow FAMA technique in \cite{wong2023sFAMA,Xu2024revisiting}. In slow FAMA, the FAS at each user only needs to switch its position once during each channel coherence period and remains unchanged unless the CSI changes. Finding the best position is also much easier as only the knowledge of the {\em average} signal-to-interference plus noise ratio (SINR) at the ports is required. There are also further efforts to exploit deep learning to assist the process of estimating the port SINR for slow FAMA \cite{Waqar-2023,Eskandari-2024}.

While slow FAMA is appealing for practical reasons, fast FAMA performs much better in terms of the number of users that can be supported \cite{Wang-aifas2024}. For example, it is possible for fast FAMA to handle hundreds of users while slow FAMA is only able to deal with several users on the same physical channel.\footnote{Note that this comment assumes a not so ambitious SINR target for each user, and the exact number of supportable users reduces greatly if the required SINR at the users increases.} To improve multi-access capability while maintaining simplicity, a variant of slow FAMA, coined as compact ultra massive array (CUMA), was proposed in \cite{Wong2024cuma}. This method shares the simplicity of slow FAMA to adapt the antenna position to only once each coherence time, but activates many `coherent' ports for analogue signal mixing. CUMA can accommodate tens of co-channel users and works well even for channels with finite scattering if randomized RISs are deployed \cite{Wong-2024cuma-ris}. In addition, CUMA at the users combines well with multiuser MIMO at the BS to reduce the burden of CSI at the BS side for massive access \cite[Section V-E]{New-2024tut}. On the other hand, recent attempts have also found channel coding to be a crucial ingredient to strengthen slow FAMA \cite{hong2024coded,hong2025Downlink}. Opportunistic scheduling is another technique to leverage slow FAMA for enhancing the network performance \cite{Waqar-2024ofama}. Interestingly, although FAMA can be a standalone multiple access scheme, FAS can be used to facilitate more reliable operations of NOMA \cite{New-2024noma} and RSMA \cite{Rostami2024fluid} for realizing their capacity advantages.

\vspace{-2mm}
\subsection{Challenges and Contributions}
In summary, CUMA and slow FAMA are practically attractive but they are not designed to handle the IUI for supporting hundreds of users or more on the same channel. This is what fast FAMA promises to do theoretically but it is not known to be practically achievable. Yet, massive access methods without the need for CSI at the BS side nor centralized optimization are profoundly important for 6G and beyond systems. In this paper, our ambition is to devise a FAMA scheme that performs better than fast FAMA and is practically realizable.

To achieve this goal, it is natural to resort to AI techniques due to the complexity and unknown nature of the problem. In fact, there is an increasing trend of employing AI to overcome complex wireless optimization problems in real time, and they are poised to play a major role in 6G \cite{Zappone-2019}. AI-driven methods to learn intricate interference patterns, make rapid data-driven decisions for resource allocation, beamforming, and decoding, overcoming the limitations of traditional model-based methods have previously been proposed \cite{Challita-2019,Shen-2021gnn,Zhang-icspcc2018}. For example, deep neural networks (DNNs) have been successfully applied to multiuser detection, demonstrating robustness to interference and channel distortions, while reinforcement learning has been used for dynamic spectrum access and power control in interference-limited networks \cite{Lohan-2024,Wu-2020,Waqar-2022}.

In the context of FAMA, AI methods are capable of learning to map observed interference metrics to optimal antenna port selections or predict future interference levels for proactive port switching. In this regard, generative AI models, such as generative adversarial networks and diffusion models, which iteratively denoise random inputs to recover clean signals, are particularly promising \cite{Letafati-2023,Wang-2025,Du-2024}. Recently, these models have been employed for channel estimation, signal detection, and constellation shaping \cite{Wu-2024,Pan-2024,Jin-2024,Fesl-2024}. Their ability to handle high-dimensional data and operate in real time after training motivates our work to integrate an AI-driven diffusion module into FAMA for enhanced interference suppression.

Another promising approach to improve resilience of wireless links is joint source-channel coding (JSCC) \cite{Lin-2024,Deniz-2024}. JSCC jointly designs the transmitted signals by taking into account both the source characteristics (for compression) and the channel conditions (for error correction), thereby introducing redundancy that mirrors the source content and significantly enhances system robustness \cite{Bourtsoulatze-2019,Kurka-2020}. In interference-prone environments, JSCC can exploit the inherent structure of both the source and the channel interference to achieve robustness unattainable by classical error-correcting codes designed for independent and identically distributed (i.i.d.) errors. For this reason, in this work, we will leverage JSCC by integrating a diffusion-based denoiser into the decoder, thereby treating the interference-plus-noise signal as a contaminant to be removed during joint detection and decoding. Instead of encoding each symbol independently, our system embeds redundancy across symbols, so that even if some symbols are heavily corrupted by interference as well as noise, the overall message can still be reliably recovered. This integration of JSCC with FAMA not only harnesses the spatial diversity of fluid antennas but further provides robust diversity, resulting in an interference-resilient communication scheme. Motivated by the limitations of current FAMA schemes and recent advances in denoising diffusion and JSCC models, we leverage AI-driven diffusion-based denoising and deep JSCC to improve FAMA. 

Specifically, in this paper, we develop a new FAMA scheme that heuristically shortlists a few promising ports that produce an output signal using maximum ratio combining (MRC). This scheme is integrated with an AI-driven interference mitigation framework and JSCC. The proposed scheme is referred to as turbo FAMA. The contributions are summarized as follows.
\begin{itemize}
\item First, we propose a new FAMA architecture that shortlists several promising ports for MRC.
\item Moreover, we develop an AI-driven interference mitigation framework based on a mix diffusion-based denoising model, which adaptively learns and suppresses residual interference which exhibits uncertain characteristics.
\item Also, we integrate a deep JSCC scheme into the FAMA framework, ensuring robust end-to-end (e2e) data transmission by jointly optimizing source and channel coding in interference-prone environments.
\item Simulation results in terms of symbol error rate (SER) will demonstrate that the proposed turbo FAMA scheme outperforms greatly fast FAMA with ideal port selection and can support an extremely large number of users on the same physical channel without CSI at the BS.
\end{itemize}

The rest of this paper is organized as follows. Section \ref{sec:system_model} introduces the system model, describing the channel model of FAS and the interference network model. Section \ref{sec:suboptimal_selection_math_extended} presents our port-selection strategies while Section \ref{sec:diffusion_denoising} details the proposed diffusion-based denoising and its integration with the deep JSCC scheme. In Section \ref{sec:simulation_results}, we provide the simulation results. Finally, Section \ref{sec:conclusion} concludes the paper.

\vspace{-2mm}
\section{System Model}\label{sec:system_model}
Consider a single-cell downlink system in which a BS with \( N_t \) spatially distributed antennas serves \( U \) single-antenna user terminals (UTs), with \( N_t = U \). Each BS antenna is dedicated to serving one UT, such that the \( u \)-th BS antenna primarily transmits data intended for UT \( u \), for \( u = 1, 2, \dots, U \). At the user side, each UT adopts an FAS comprising $K$ discrete ports arranged linearly along an aperture of length \( W\lambda \), where \( \lambda \) denotes the carrier wavelength.\footnote{This work can be easily extended to the case if each UT is equipped with a two-dimensional (2D) FAS.} The FAS enables fast position switching of the active port among the $K$ positions so that it is reasonable to assume that with FAS, UT $u$ can access the received signals from all $K$ positions in the given space.\footnote{This might mean that each UT switches its port $K$ times in each symbol duration, or $\frac{K}{n_{\rm RF}}$ times if it has $n_{\rm RF}$ radio-frequency (RF) chains. Evidently, it is also possible to reduce the number of switching times per symbol duration by exploiting the spatial correlation between the positions using AI techniques, which was studied in \cite{Wong-vfas2024}. However, we defer a full investigation of how this technique works in the context of FAMA to future work.}

\vspace{-2mm}
\subsection{Signal Model}
Let \( s_u \in \mathbb{C} \) be the transmitted symbol from the \( u \)-th BS antenna, satisfying \( \mathbb{E}[|s_u|^2] = \sigma_s^2 \) and assumed independent across \( u \). The complex channel gain from the \( \bar{u} \)-th BS antenna to the \( k \)-th port of user \( u \) is denoted by \( g^{(\bar{u},u)}_{k} \). The received baseband signal at the \( k \)-th port of UT \( u \) is expressed as
\begin{equation}\label{eq:rx_signal_sectionII}
r^{(u)}_k = g^{(u,u)}_{k} s_u + \sum_{\bar{u} = 1\atop \bar{u} \neq u}^{U} g^{(\bar{u},u)}_{k} s_{\bar{u}} + \eta^{(u)}_k,
\end{equation}
where \( \eta^{(u)}_k \sim \mathcal{CN}(0, \sigma_\eta^2) \) is the circularly symmetric complex Gaussian (CSCG) noise at port \( k \) of UT \( u \), independent across \( u \) and \( k \). The average received signal-to-noise ratio (SNR) per port for the desired signal is defined as
\begin{equation}\label{eq:snr_def}
\Gamma_u \triangleq \frac{\Omega_u \sigma_s^2}{\sigma_\eta^2},
\end{equation}
where \(\Omega_u = \mathbb{E}[|g^{(u,u)}_{k}|^2]\) is the average channel gain of the desired signal, assumed constant across ports \( k \) for UT \( u \).


\vspace{-2mm}
\subsection{Channel Model}
The value of $K$ is usually large and hence the ports in FAS are closely spaced, resulting in high spatial correlation among the channel coefficients within the length of \( W\lambda \). For UT \( u \), define the channel vector from the \( u \)-th BS antenna to the \( K \) ports as \( \bm{g}_{u,u} = [g^{(u,u)}_{1}, g^{(u,u)}_{2}, \dots, g^{(u,u)}_{K}]^T \). We assume that \( \bm{g}_{u,u} \sim \mathcal{CN}(\bm{0}, \Omega_{u,u} \bm{J}) \), where \( \Omega_{u,u} = \mathbb{E}[|g^{(u,u)}_{k}|^2] \) is the average channel gain from BS antenna \( u \) to UT \( u \)'s ports, and \( \bm{J} \in \mathbb{C}^{K \times K} \) is the spatial correlation matrix. The channel vectors \( \bm{g}_{u,u} \) are assumed independent across different \( u \). Also, the \( (n,m) \)-th element of \( \bm{J} \) is given by
\begin{equation}\label{eq:covariance_matrix}
[\bm{J}]_{n,m} = J_0\left(2\pi \frac{|n - m|}{K - 1} W\right),
\end{equation}
in which \( J_0(\cdot) \) is the zeroth-order Bessel function of the first kind, modeling correlation as a function of port separation.

To generate \( \bm{g}_{u,u} \), we employ the eigenvalue decomposition of \( \bm{J} = \bm{Q} \bm{\Lambda} \bm{Q}^H \), where \( \bm{Q} \) is a unitary matrix of eigenvectors, and \( \bm{\Lambda} = {\rm diag}(\lambda_1, \lambda_2, \dots, \lambda_K) \) contains the eigenvalues satisfying \( \lambda_1 \geq \lambda_2 \geq \cdots \geq \lambda_K \geq 0 \). Thus,
\begin{equation}\label{eq:channel_generation}
\bm{g}_{u,u} = \sqrt{\Omega_{u,u}} \bm{Q} \bm{\Lambda}^\frac{1}{2} \bm{w}_{v,u},
\end{equation}
in which \( \bm{w}_{u,u} \sim \mathcal{CN}(\bm{0}, \bm{I}_K) \) is a vector of independent CSCG entries. This ensures \( \mathbb{E}[\bm{g}_{u,u} \bm{g}_{u,u}^H] = \Omega_{u,u} \bm{J} \), preserving the desired correlation structure.

\vspace{-2mm}
\subsection{Symbol Level Port Switching}
Fast FAMA leverages symbol-level port selection to maximize the instantaneous SINR. In particular, fast FAMA adapts the active port \( k_u^* \) for each symbol based on real-time channel conditions in the data-dependent interference-plus-noise signal. Define the instantaneous SINR at port \( k \) of UT \( u \) as
\begin{equation}\label{eq:sinr_instantaneous}
\gamma^{(u)}_k = \frac{|g^{(u,u)}_{k}|^2 |s_u|^2}{\left| \sum_{\bar{u} \neq u} g^{(\bar{u},u)}_{k} s_{\bar{u}} + \eta^{(u)}_k \right|^2}.
\end{equation}
Ideally, UT \( u \) should select
\begin{equation}\label{eq:fast_fama_ideal}
k_u^* = \arg \max_{1 \leq k \leq K} \gamma^{(u)}_k.
\end{equation}
Since \( |s_u|^2 \) is constant across ports for a given symbol, this is equivalent to
\begin{equation}\label{eq:fast_fama_ratio}
k_u^* = \arg \max_{1 \leq k \leq K} \frac{|g^{(u,u)}_{k}|^2}{\left| \sum_{\bar{u} \neq u} g^{(\bar{u},u)}_{k} s_{\bar{u}} + \eta^{(u)}_k \right|^2}.
\end{equation}

The efficacy of fast FAMA arises from exploiting deep fades in the data-dependent interference term \( \sum_{\bar{u} \neq u} g^{(\bar{u},u)}_{k} s_{\bar{u}}+ \eta^{(u)}_k \). For large \( U \), central limit theorem (CLT) applies and this term behaves like a complex Gaussian random variable. As a result, the magnitude \( \left| \sum_{\bar{u} \neq u} g^{(\bar{u},u)}_{k} s_{\bar{u}}+ \eta^{(u)}_k \right| \) is Rayleigh distributed, which is well known to suffer from deep fades with significant likelihood. These deep fades enable the fast FAMA mechanism to choose a port \( k \) in which the aggregate interference signal is momentarily suppressed, boosting the instantaneous SINR.

Despite its effectiveness to neutralize interference by natural fading, estimation of the instantaneous, data-dependent SINR for port selection is non-trivial. This issue was tackled in \cite{Wong-2024ffama} but considerable performance degradation was observed.


\vspace{-2mm}
\section{Heuristics-based Port Shortlisting}\label{sec:suboptimal_selection_math_extended}
Accurately estimating the instantaneous SINR at the ports to determine the best port is very challenging. Here, we present several heuristics that identify ports with desirable properties for processing. The heuristics are simple and realizable.


\vspace{-2mm}
\subsection{Port Shortlisting}\label{subsec:mdsp_nss_sdm_math}
Given that each user, say UT $u$, possesses the knowledge of the desired channel \( g^{(u,u)}_{k} \) \cite{xu2024channel,zhang2023successive,Xu-2025ce,10751774}, the predicted received power of the desired signal can be expressed as
\begin{equation}\label{eq:desired_power}
P_{\text{desired}}^{(u)}(k)=| g^{(u,u)}_{k}|^2 \sigma_s^2.
\end{equation}
A straightforward suboptimal selection rule involves identifying those ports \( k \) whose power \( |r^{(u)}_k|^2 \) most closely matches \eqref{eq:desired_power}. Large deviations from this value suggest either a strong interference component or destructive fading of the desired signal. As such, we define the deviation metric as
\begin{equation}\label{eq:deviation_metric}
\Delta^{(u)}_k \triangleq \left| |r^{(u)}_k|^2 - P_{\text{desired}}^{(u)}(k) \right|.
\end{equation}
Ports are then ranked in ascending order of \( \Delta^{(u)}_k \). 

To quantify the reliability of \eqref{eq:deviation_metric}, consider the received signal \( r^{(u)}_k = g^{(u,u)}_{k} s_u + I^{(u)}_k + \eta^{(u)}_k \), where \( I^{(u)}_k \triangleq \sum_{\bar{u} \neq u} g^{(\bar{u},u)}_{k} s_{\bar{u}} \). For large \( U \), \( I^{(u)}_k \) approximates a complex Gaussian random variable by CLT, with zero mean and approximate variance of \( \sigma_I^2 = (U-1)\sigma_s^2 \Omega_{\text{cross}} \), where \( \Omega_{\text{cross}} = \mathbb{E}[| g^{(\bar{u},u)}_{k}|^2] \). The received power \(| r^{(u)}_k|^2 \) follows a noncentral chi-squared distribution with $2$ DoF, parameterized by the desired signal power \(| g^{(u,u)}_{k}|^2 \sigma_s^2 \) and the interference-plus-noise variance \( \sigma_I^2 + \sigma_\eta^2 \). The probability that \( | r^{(u)}_k|^2 \) deviates from \( P_{\text{desired}}^{(u)}(k) \) by less than \( \Delta \) can be bounded by using
\begin{equation}
\mathbb{P}\left( \Delta^{(u)}_k<\Delta \right) = 1-\mathcal{F}_{\chi^2}\left( \frac{\Delta}{\sigma_I^2 + \sigma_\eta^2}; 2, \frac{P_{\text{desired}}^{(u)}(k)}{\sigma_I^2 + \sigma_\eta^2} \right),
\end{equation}
in which \( \mathcal{F}_{\chi^2} \) denotes the noncentral chi-squared cumulative density function (CDF). This bound provides some theoretical insight for selecting ports with minimal deviation.

To refine this approach, we normalize the deviation by the predicted desired-signal power, yielding
\begin{equation}\label{eq:port_selection_normalized}
\Delta^{(u)}_k=\frac{ \Bigl| \bigl| r^{(u)}_k \bigr|^2 - \bigl| g^{(u,u)}_{k} \bigr|^2 \sigma_s^2 \Bigr| }{ \bigl| g^{(u,u)}_{k} \bigr|^2 \sigma_s^2 },
\end{equation}
where the relative difference governs the decision. This normalization prevents ports with strong desired channels from being unfairly penalized due to larger absolute deviations.

Additionally, in systems with numerous ports, pre-filtering weak desired-channel gains improves efficiency. To do so, we define the maximum desired-channel gain as
\begin{equation}\label{eq:max_gain}
\bigl| g^{(u,u)}_{\max} \bigr|^2\triangleq\max_{1 \leq k \leq K} \bigl| g^{(u,u)}_{k} \bigr|^2,
\end{equation}
and select a candidate set
\begin{equation}\label{eq:mdsp_def}
\mathcal{K}^{(u)} = \left\{ k: \bigl| g^{(u,u)}_{k} \bigr|^2 \geq \gamma_{\text{th}} \bigl| g^{(u,u)}_{\max}\bigr|^2 \right\},
\end{equation}
where \( 0 < \gamma_{\text{th}} < 1 \) is a threshold parameter. This set excludes ports with negligible desired signals, reducing the risk of selecting interference-dominated ports.

\begin{figure*}
\centering
\begin{multicols}{3}
\includegraphics[width=2.5in]{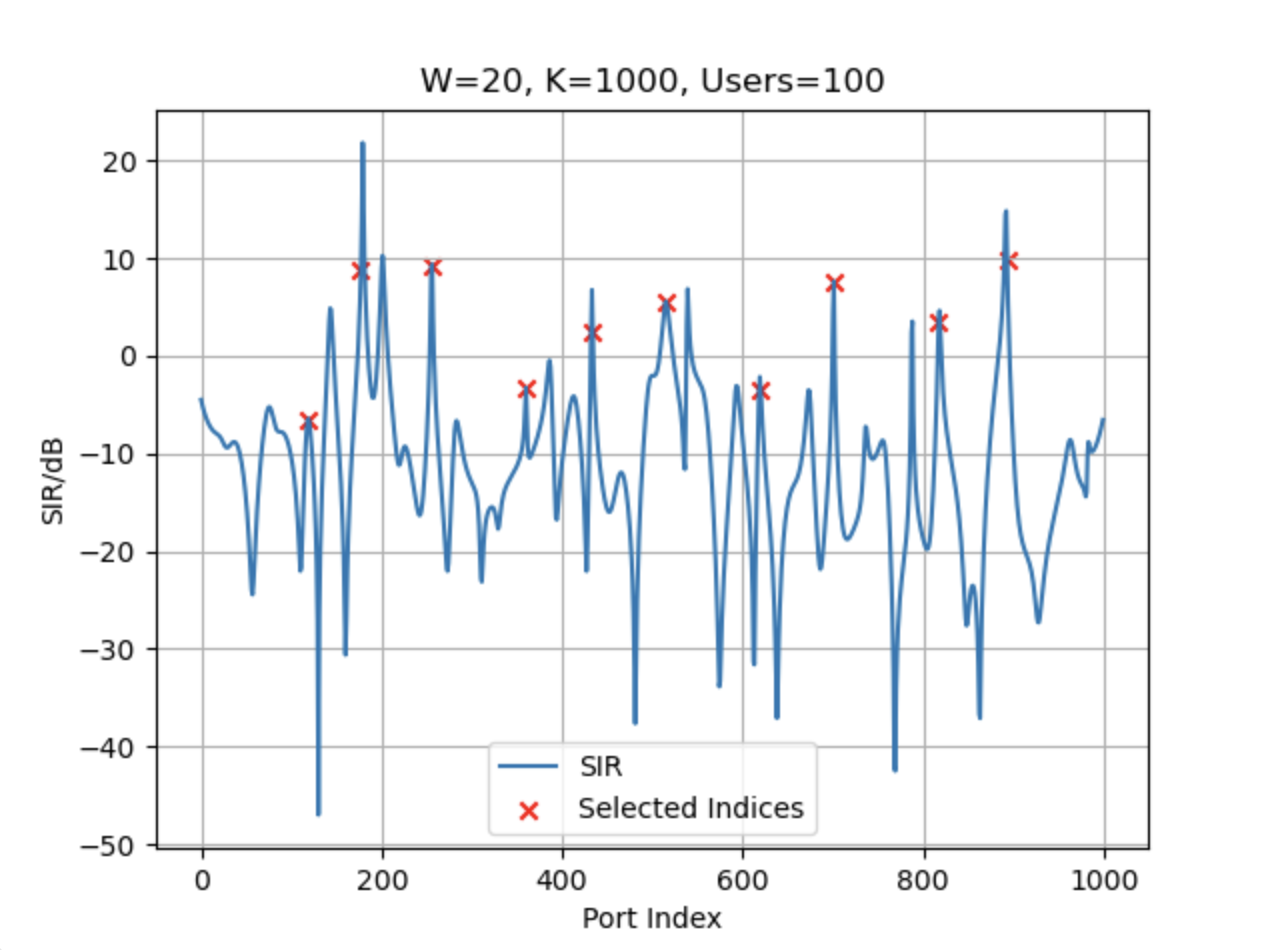}\\
\par
\includegraphics[width=2.5in]{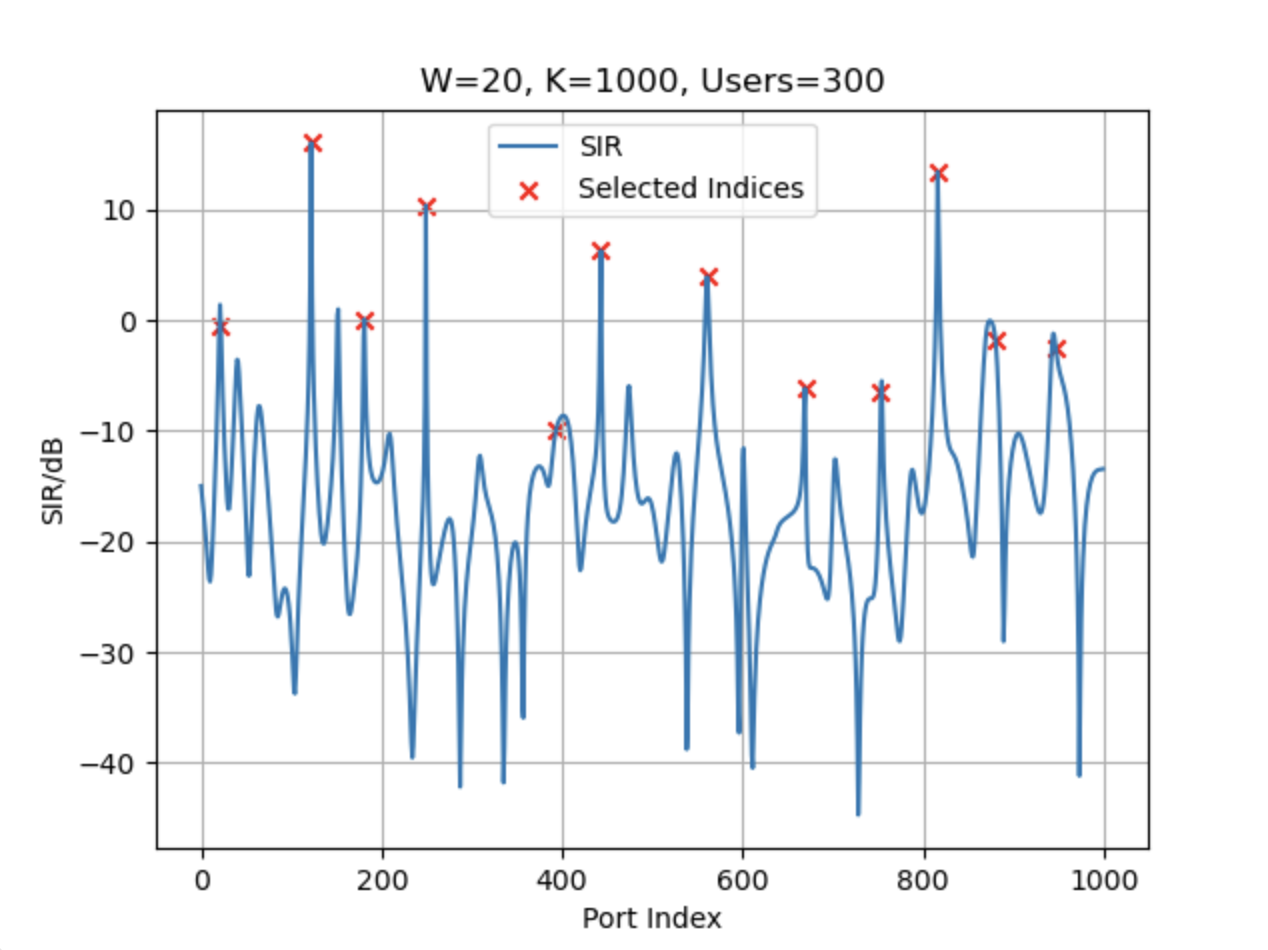}\\
\includegraphics[width=2.5in]{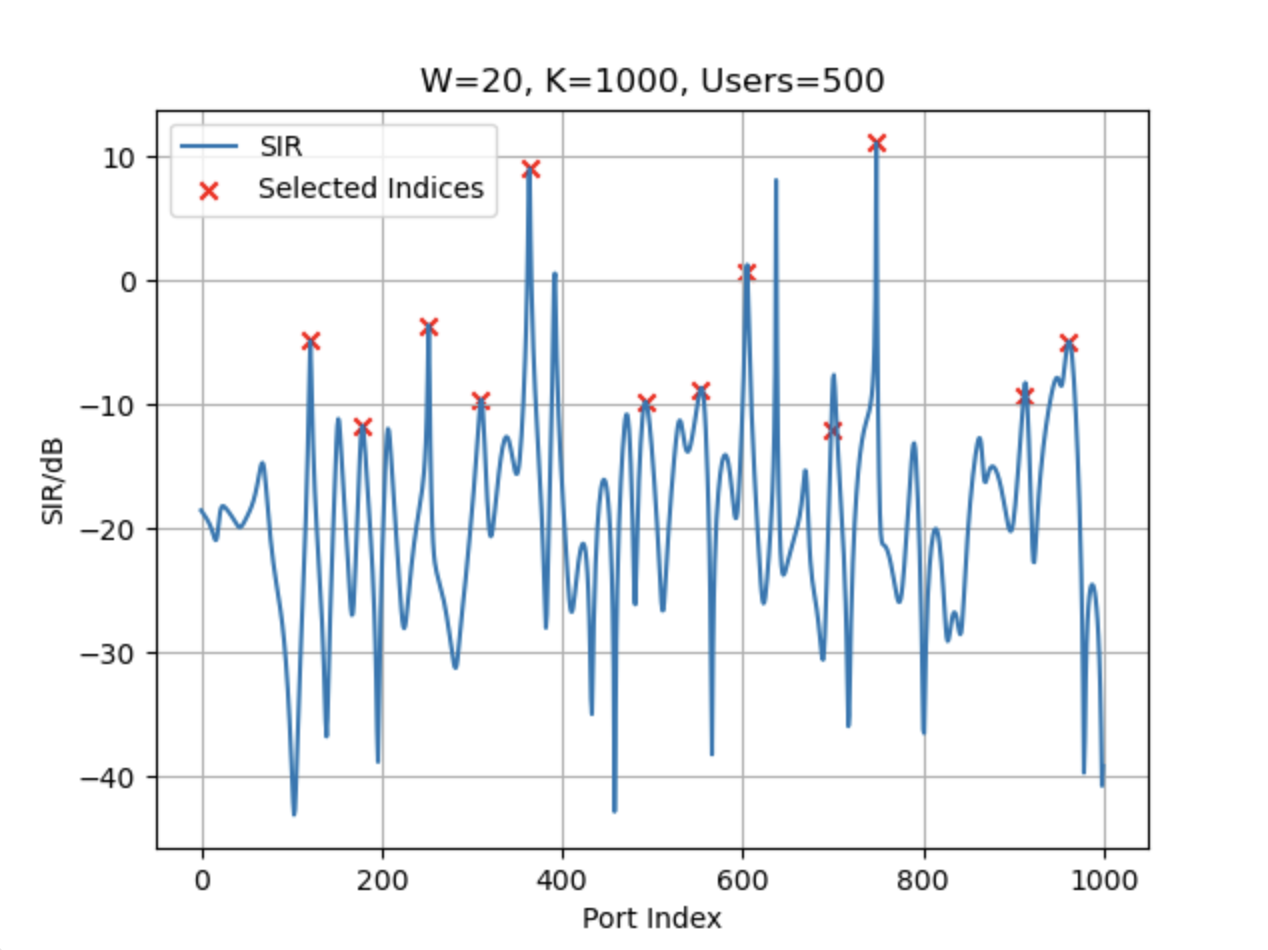}
\end{multicols}
\vspace{-15pt}
\caption{Illustration of the port shortlisting process with $\gamma_\text{th}=0.6$, $K=1000$ ports and $W=20$, for $100$, $300$, and $500$ users. The blue lines represent the instantaneous SINR across all port indices, while the red `$\times$' markers indicate the selected ports that satisfy the shortlisting criteria (without SDM).}\label{fig:shortlisting}
\vspace{-15pt}
\end{figure*}

The third rule is regarding spatial diversity maximization (SDM). From \( \mathcal{K}^{(u)} \), we select \( K_{\text{sel}} \) ports to maximize spatial diversity within the aperture \( W\lambda \). The optimal set \( \mathcal{K}_{\text{SDM}}^{(u)} \) solves the following combinatorial optimization:
\begin{equation}\label{eq:sdm_maxmin}
\mathcal{K}_{\text{SDM}}^{(u)} = \arg \max_{\mathcal{K} \subset \mathcal{K}^{(u)}\atop |\mathcal{K}| = K_{\text{sel}}} \min_{k,\ell \in \mathcal{K}} \delta(k,\ell),
\end{equation}
where the normalized port separation is given by
\begin{equation}\label{eq:port_separation}
\delta(k,\ell) = \frac{|k -\ell|}{K-1} W.
\end{equation}
This max-min formulation ensures that the selected ports are as far apart as possible, decorrelating their channels for more diversity benefits. This selection process is illustrated in Fig.~\ref{fig:shortlisting} in which SDM is not applied but we impose a fixed minimum spacing between adjacent selected ports $d\times K$ with $d=0.05$. 

Overall, the heuristic-based port shortlisting scheme considering all the procedures above is given in Algorithm \ref{alg:port_selection}.

\begin{algorithm}[H]
\caption{Port Shortlisting Process}\label{alg:port_selection}
\begin{algorithmic}[1]
\Require Knowledge of CSI \( \{ g^{(u,u)}_{k} \} \), threshold \( \gamma_{\text{th}} \)
\Ensure Selected port set \( \mathcal{K}_{\text{SDM}}^{(u)} \)
\State Compute predicted desired power for all ports using \eqref{eq:desired_power}
\State Compute deviation metric \( \Delta^{(u)}_k \) for all ports using \eqref{eq:port_selection_normalized}
\State Rank ports in ascending order of \( \Delta^{(u)}_k \)
\State Apply desired power filtering to the ranked list using \eqref{eq:mdsp_def}
\State Solve the SDM from the filtered list using \eqref{eq:sdm_maxmin}
\end{algorithmic}
\end{algorithm}

\vspace{-2mm}
\subsection{MRC for Shortlisted Ports}\label{subsec:mrc_implementation}
After selecting \( K_{\text{sel}} \) ports using Algorithm \ref{alg:port_selection}, MRC is used to coherently combine the signals for maximizing the effective SNR. Specifically, MRC weights the signal of each shortlisted port by the conjugate of its normalized channel gain to produce the output signal for UT $u$ as
\begin{equation}\label{eq:mrc_combine}
y_u 
= \sum_{k \in \mathcal{K}_{\text{SDM}}^{(u)}} \frac{ \left( g^{(u,u)}_{k} \right)^* }{ \sqrt{ \Omega_u } } r^{(u)}_k.
\end{equation}

MRC introduces minimal complexity, without requiring any knowledge of the interference. It is worth stating that MRC is performed in the computational domain rather than in the signal domain, meaning that using FAS, fast port switching per symbol period allows for the received signals to be obtained on one or a small number of RF chains. Once all the received signals $\{r^{(u)}_k\}$ are obtained, they are recorded and processed in the computational domain to produce $y_n$ for decoding.

\vspace{-2mm}
\section{Diffusion-based Denoising for Residual Interference Removal}\label{sec:diffusion_denoising}
The proposed heuristics mitigate dominant interference by shortlisting ports with favorable conditions based on the set criteria. However, the residual interference plus noise at the combined signal output after MRC, denoted as \( \eta_{\text{res}}\), remains. This residual interference exhibits non-Gaussian characteristics such as heavier tails and multi-modal distributions due to partial constructive alginment or destructive misalignment of multiuser signals. Classical detectors, such as linear or MMSE estimators, rely on Gaussian noise assumptions and thus incur significant performance degradation under these conditions. To overcome this, here, we propose a diffusion-based denoising framework that leverages generative modeling to learn and remove structured residual interference and noise, enhancing symbol detection accuracy in the FAMA system.

The motivation behind adopting diffusion-based denoising stems from the dynamic and highly sophisticated nature of this data-dependent residual interference, $\eta_{\text{res}}$. Symbol-level port switching introduces chaotic, but correlated variations in interference patterns, rendering traditional model-based approaches requiring precise CSI or multiuser inversion infeasible. On the contrary, diffusion models, operating in a data-driven manner, adaptively capture the statistical distribution of clean signals and interference without explicit CSI, thereby accommodating non-Gaussianity and multi-modality. 


\vspace{-2mm}
\subsection{Denoising Diffusion Probabilistic Models (DDPM)}\label{subsec:ddpm_review}
DDPM \cite{Ho2020} employs a two-phase process, a \emph{forward diffusion} phase that corrupts a clean signal with noise over \( T \) steps, and a \emph{reverse denoising} phase that reconstructs the original signal from its noisy version. It should be noted that $T$ steps here may mean the number of iterations required for training the model or that for iterative processing when applied.

\subsubsection{Forward diffusion process}  
Let \( \mathbf{x}_0 \in \mathbb{C}^n \) represent the clean received symbol vector after MRC in which $n$ denotes the number of signal samples in time (i.e., $y_u(1),\dots,y_u(n)$). For mathematical convenience, we stack its real and imaginary parts, treating \( \mathbf{x}_0 \in \mathbb{R}^{2n} \). Define a noise schedule \( \{\beta_t\}_{t=1}^T \), where \( \beta_t \in (0,1) \), \( \alpha_t = 1 - \beta_t \), and \( \bar{\alpha}_t = \prod_{i=1}^t \alpha_i \). The forward diffusion process is performed by
\begin{equation}\label{eq:forward_process}
q(\mathbf{x}_t|\mathbf{x}_{t-1}) = \mathcal{N}\left( \mathbf{x}_t; \sqrt{\alpha_t} \mathbf{x}_{t-1}, \beta_t \mathbf{I} \right),~\mbox{for } 1 \leq t \leq T,
\end{equation}
with the joint distribution factorized as
\begin{equation}\label{eq:forward_factorization}
q({\bf x}_{1:T}|\mathbf{x}_0) = \prod_{t=1}^T q(\mathbf{x}_t|\mathbf{x}_{t-1}),
\end{equation}
where ${\bf x}_{1:T}$ is the shorthand for $\mathbf{x}_1,\mathbf{x}_2,\dots,\mathbf{x}_T$. This process incrementally adds Gaussian noise, scaling the signal by \( \sqrt{\alpha_t} \) and introducing variance \( \beta_t \), to allow direct sampling
\begin{equation}\label{eq:direct_sample}
\mathbf{x}_t = \sqrt{\bar{\alpha}_t} \mathbf{x}_0 + \sqrt{1 - \bar{\alpha}_t} \boldsymbol{\epsilon},~\mbox{where } \boldsymbol{\epsilon} \sim \mathcal{N}(\mathbf{0}, \mathbf{I}).
\end{equation}
After \( T \) steps, we have \( \mathbf{x}_T \approx \mathcal{N}(\mathbf{0}, \mathbf{I}) \), effectively erasing the original signal's structure and containing only noise.

\subsubsection{Reverse denoising process}  
The reverse process, parameterized by \( p_\theta \), approximates the posterior
\begin{equation}\label{eq:reverse_gaussian}
p_\theta(\mathbf{x}_{t-1}|\mathbf{x}_t) = \mathcal{N}\left( \mathbf{x}_{t-1}; \boldsymbol{\mu}_\theta(\mathbf{x}_t), \sigma_t^2 \mathbf{I} \right),
\end{equation}
where \( \boldsymbol{\mu}_\theta(\mathbf{x}_t) \) is predicted by a neural network, and \( \sigma_t^2 \) is either fixed or learned. Typically, we have
\begin{equation}\label{eq:mu_theta}
\boldsymbol{\mu}_\theta(\mathbf{x}_t) = \frac{1}{\sqrt{\alpha_t}} \left( \mathbf{x}_t - \frac{\beta_t}{\sqrt{1 - \bar{\alpha}_t}} \boldsymbol{\epsilon}_\theta(\mathbf{x}_t) \right),
\end{equation}
with \( \boldsymbol{\epsilon}_\theta(\mathbf{x}_t) \) estimating the noise \( \boldsymbol{\epsilon} \).

\subsubsection{Training objective}  
Training minimizes a simplified variational bound, reducing to noise prediction
\begin{equation}\label{eq:DDPM_loss}
\mathcal{L}_{\text{DDPM}} = \mathbb{E}_{t, \mathbf{x}_0, \boldsymbol{\epsilon}} \left[ \left\| \boldsymbol{\epsilon} - \boldsymbol{\epsilon}_\theta(\mathbf{x}_t) \right\|^2 \right].
\end{equation}
Sampling involves
\begin{equation}\label{eq:reverse_sampling}
\left\{\begin{aligned}
\mathbf{x}_T &\sim \mathcal{N}(\mathbf{0}, \mathbf{I}),\\
\mathbf{x}_{t-1} &\sim p_\theta(\mathbf{x}_{t-1}|\mathbf{x}_t),~\mbox{for } t = T, \dots, 1.
\end{aligned}\right.
\end{equation}
In symbol-level FAMA (e.g., fast FAMA), \( \mathbf{x}_0 \) represents the desired signal, and \( \eta_{\text{res}} \) is treated as the initial corruption. But DDPM Gaussian noise assumption limits its ability to model multi-modal interference, prompting an extension.

\subsection{Mixture-based Denoising Diffusion (MixDDPM)}\label{subsec:mixddpm}
\begin{figure}
\begin{center}
\includegraphics[width=\columnwidth]{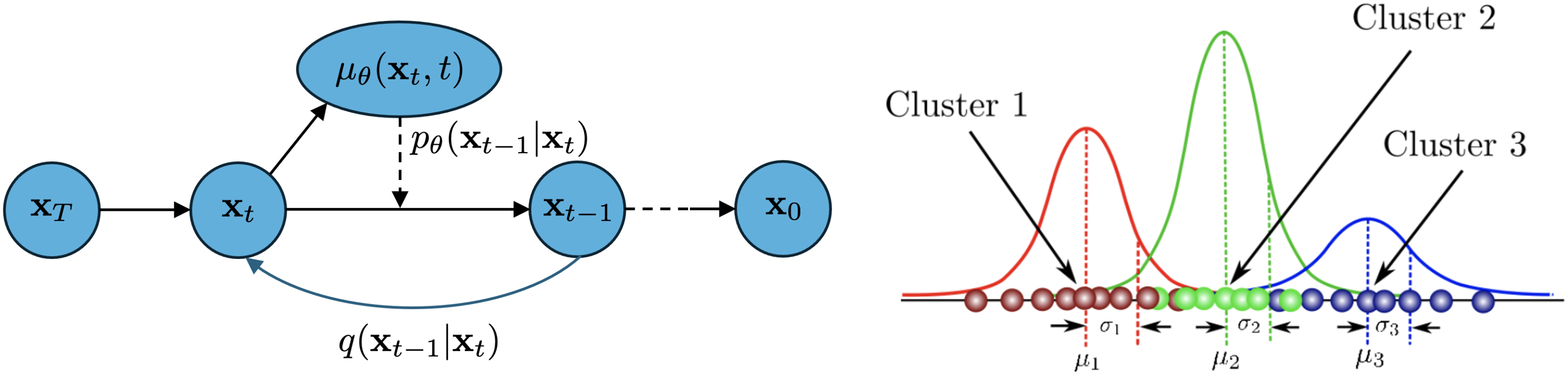}
\caption{The MixDDPM framework, depicting the forward diffusion process $q(\mathbf{x}_{t}|\mathbf{x}_{t-1})$ that gradually corrupts the clean signal and the reverse denoising process $p_\theta(\mathbf{x}_{t-1}|\mathbf{x}_{t})$ that iteratively recovers $\mathbf{x}_0$.}\label{fig:mixddpm}
\vspace{-4mm}
\end{center}
\end{figure}

To address the limitations of standard DDPMs in modeling the complex, multi-modal interference distributions encountered in symbol-level FAMA systems, we introduce the MixDDPM, as shown in Fig.~\ref{fig:mixddpm} \cite{Nachmani-2021,Guo-2023}. This framework extends the traditional model by incorporating Gaussian mixture noise, enabling explicit representation of interference modes (e.g., constructive/destructive alignment) in fast FAMA systems. The key innovation lies in the ability to capture the structured nature of residual interference, which exhibits non-Gaussian characteristics due to the superposition of multiuser signals.

\subsubsection{Forward process with mixture noise}  
The forward diffusion process in MixDDPM is defined as
\begin{equation}\label{eq:mix_forward}
q(\mathbf{x}_t|\mathbf{x}_{t-1}) = \sum_{j=1}^J \omega_j \mathcal{N}\left( \mathbf{x}_t; \sqrt{\alpha_t} \mathbf{x}_{t-1} + \sqrt{\beta_t} \mathbf{c}_j, \beta_t \mathbf{I} \right),
\end{equation}
in which \( \{\mathbf{c}_j\}_{j=1}^J \) are the mixture centers, \( \omega_j \) are the mixture weights satisfying \( \sum_{j=1}^{K} \omega_j = 1\), and $J$ denotes the number of modes. All of these parameters can be fixed or learned. The marginal distribution at step \( t \) follows (\ref{eq:direct_sample}) as before.
 
The Gaussian-mixture diffusion process allows for closed-form sampling of \( \mathbf{x}_t \) at any timestep \( t \). To do so, let \( \gamma_{t,i} = \beta_i \prod_{j=i+1}^{t} \alpha_j \) for \( i \in [1, t-1] \) and \( \gamma_{t,t} = \beta_t \). Considering the iterates of \( \mathbf{x}_t \), we then have
\begin{equation}\label{eq:mix_iterates}
\mathbf{x}_t = \sqrt{\bar{\alpha}_t} \mathbf{x}_0 + \sqrt{\sum_{j=1}^{J} \gamma_{t,j} \sigma_j^2} \, \bar{\boldsymbol{z}}_t + \sum_{j=1}^{J} \sqrt{\gamma_{t,j}} \, \mathbf{c}_j,
\end{equation}
where \( \bar{\boldsymbol{z}}_t \sim \mathcal{N}(\mathbf{0}, \mathbf{I}) \) and \( j\) denotes the Gaussian component selected at timestep \( t \). In the sequel, when required, we will use $i_t$ to replace $j$ to highlight the time dependence. This closed-form expression simplifies the sampling process and enables efficient implementation. This is described in Algorithm \ref{alg:mixddpm_forward}.

\begin{algorithm}[H]
\caption{MixDDPM Forward Diffusion Process}\label{alg:mixddpm_forward}
\begin{algorithmic}[1]
\Require Input signal \( \mathbf{x}_0 \), noise schedule \( \{\beta_t\} \), mixture components \( \{\mathbf{c}_j, \omega_j\} \), timesteps \( T \)
\Ensure Noisy signal \( \mathbf{x}_T \)
\For{$t = 1$~to~$T$}
    \State Sample index \( j \sim \text{Categorical}(\omega_1, \dots, \omega_K) \)
    \State Sample noise \( \boldsymbol{\epsilon} \sim \mathcal{N}(\mathbf{0}, \mathbf{I}) \)
    \State Update signal by (\ref{eq:mix_iterates})
\EndFor
\end{algorithmic}
\end{algorithm}

\subsubsection{Reverse process with mixture components}  
The reverse denoising process is parameterized by
\begin{multline}\label{eq:mix_reverse}
p_\theta(\mathbf{x}_{t-1}|\mathbf{x}_t)=\\
\sum_{j=1}^{K} p_\theta(j|\mathbf{x}_t)\,\mathcal{N}\left(\mathbf{x}_{t-1};\,\boldsymbol{\mu}_\theta^{(j)}(\mathbf{x}_t),\boldsymbol{\Sigma}_\theta^{(j)}(\mathbf{x}_t)\right),
\end{multline}
where \( p_\theta(j |\mathbf{x}_t) \) classifies the interference mode, and \( \boldsymbol{\mu}_\theta^{(j)} \) and \( \boldsymbol{\Sigma}_\theta^{(j)} \) are learned parameters for each component. The mean prediction incorporates the mixture centers
\begin{equation}\label{eq:mix_mean}
\boldsymbol{\mu}_\theta^{(j)}(\mathbf{x}_t) = \frac{1}{\sqrt{\alpha_t}} \left( \mathbf{x}_t - \frac{\beta_t}{\sqrt{1 - \bar{\alpha}_t}} \boldsymbol{\epsilon}_\theta^{(j)}(\mathbf{x}_t) \right) + \mathbf{c}_j,
\end{equation}
where \( \boldsymbol{\epsilon}_\theta^{(j)}(\mathbf{x}_t) \) estimates the noise specific to component \( j \). This is summarized in Algorithm \ref{alg:mixddpm_reverse}. 

\begin{algorithm}[H]
\caption{MixDDPM Reverse Denoising Process}\label{alg:mixddpm_reverse}
\begin{algorithmic}[1]
\Require Noisy signal \( \mathbf{x}_T \), trained model \( p_\theta \), timesteps \( T \)
\Ensure Denoised signal \( \hat{\mathbf{x}}_0 \)
\State Initialize \( \mathbf{x}_{T} \)
\For{$t = T$~downto~$1$}
    \State Predict noise \( \boldsymbol{\epsilon}_\theta^{(j)}(\mathbf{x}_t, t) \) for all components \( j \)
    \State Compute mixture weights \( p_\theta(j|\mathbf{x}_t) \)
    \State Sample component index \( j \sim p_\theta(j |\mathbf{x}_t) \)
    \State Update signal by $\mathbf{x}_{t-1} =\boldsymbol{\mu}_\theta^{(j)}(\mathbf{x}_t)$ in (\ref{eq:mix_mean})
\EndFor
\end{algorithmic}
\end{algorithm}

\subsubsection{Training objective and Kullback-Leibler (KL) divergence}  
The loss function combines noise prediction and interference mode classification, given by 
\begin{equation}\label{eq:mix_loss}
\mathcal{L}_{\text{MixDDPM}} = \mathbb{E}_{t, \mathbf{x}_0, \boldsymbol{\epsilon}, j} \left[ -\log p_\theta(j | \mathbf{x}_t) + \left\| \boldsymbol{\epsilon} - \boldsymbol{\epsilon}_\theta^{(j)}(\mathbf{x}_t) \right\|^2 \right],
\end{equation}
where \( j \) is sampled from the categorical distribution defined by \( \omega_j \). The training procedure minimizes the variational lower bound (VLB) of the negative log-likelihood \cite{Wang-2024ddpm}
\begin{multline}\label{eq:mix_vlb}
L_{\text{VLB}} =-\log p_\theta(\mathbf{x}_0|\mathbf{x}_1)\\
+ \sum_{t>1} \mathbb{E}_{i_{1:t-1}|\mathbf{x}_t, \mathbf{x}_0} \left[D_{\text{KL}}\left( q(i_t | \mathbf{x}_t, \mathbf{x}_0, i_{1:t-1}) \parallel 
p_\theta(i_t | \mathbf{x}_t) \right)\right]\\
+ \sum_{t>1} \mathbb{E}_{i_{1:t}| \mathbf{x}_t, \mathbf{x}_0} \left[D_{\text{KL}}\left( q(\mathbf{x}_{t-1} | \mathbf{x}_t, \mathbf{x}_0, i_{1:t}) \parallel 
p_\theta(\mathbf{x}_{t-1} |\mathbf{x}_t, i_t) \right)\right]\\
+ D_{\text{KL}}\left( q(\mathbf{x}_T | \mathbf{x}_0) \parallel p(\mathbf{x}_T) \right),
\end{multline}
in which $D_{\text{KL}}(P\parallel Q)$ represents the KL divergence. The KL divergence terms can be computed in closed form as
\begin{align}
&\mathbb{E}_{i_{1:t} | \mathbf{x}_t, \mathbf{x}_0} \left[ D_{\text{KL}}\left( q(\mathbf{x}_{t-1} | \mathbf{x}_t, \mathbf{x}_0, i_{1:t}) \parallel p_\theta(\mathbf{x}_{t-1} | \mathbf{x}_t, i_t) \right) \right]\notag\\
=& \mathbb{E}_{i_t | \mathbf{x}_t, \mathbf{x}_0} \left[\begin{array}{l}
\frac{\sigma_t^2}{2\,\sigma_t^2(i_t)}+ \frac{1}{2\,\sigma_t^2(i_t)}\times\\
\left\|\begin{array}{l}
\mathbb{E}_{i_{1:t-1} | \mathbf{x}_t, \mathbf{x}_0, i_t} \left[ \boldsymbol{\mu}_\theta(\mathbf{x}_t,\mathbf{x}_0,i_{1:t} | i_{1:t-1}) \right]\\
-\boldsymbol{\mu}_\theta(\mathbf{x}_t, t, i_t)
\end{array}\right\|_2^2\\
+ \log \frac{\sigma_t(i_t)}{\sigma_t}
\end{array}\right]\notag\\ 
&\quad\quad\quad\quad\quad + C,\label{eq:mix_kl}
\end{align}
where \( C \) is a constant independent of \( \theta \).
  
To predict \( \boldsymbol{\mu}_\theta(\mathbf{x}_t) \), we use the parameterization
\begin{equation}\label{eq:mix_param}
\boldsymbol{\mu}_\theta(\mathbf{x}_t,i_t) = \frac{1}{\sqrt{\bar{\alpha}_t}} \left( \mathbf{x}_t - \sqrt{\beta_t} \mathbf{c}_{i_t} - \frac{\beta_t}{\sqrt{1 - \bar{\alpha}_t}} \boldsymbol{\epsilon}_\theta(\mathbf{x}_t) \right),
\end{equation}
where \( \boldsymbol{\epsilon}_\theta(\mathbf{x}_t) \) denotes a neural network predicting the noise component. This parameterization ensures that the model can effectively learn the structure of interference modes while maintaining computational efficiency.
 
The Gaussian-mixture diffusion process permits simple, closed-form sampling of \( \mathbf{x}_t \) at any timestep \( t \). Furthermore, the use of fast ordinary differential equation (ODE) solvers during the reverse process accelerates convergence, making the framework suitable for real-time applications in symbol-level FAMA systems. The explicit modeling of interference modes through mixture components enhances the model's ability to adapt to fast changing channel conditions.

In the proposed FAMA, MixDDPM addresses the challenge of residual interference by explicitly modeling the multi-modal nature of interference. The mixture components correspond to different interference alignment scenarios (e.g., constructive versus destructive), enabling the denoiser to adaptively remove structured interference. This integration improves symbol detection accuracy in interference-limited regimes, a key advantage over traditional Gaussian-based denoising methods.

\vspace{-2mm}
\subsection{JSCC with Denoising Diffusion}\label{sec:JSCC_diffusion}
To enhance the robustness of the proposed FAMA system, we combine the diffusion-based denoising pipeline with a deep JSCC framework (encoder and decoder). The JSCC encoder transforms digital data symbols such as quaternary phase shift keying (QPSK) into a latent space representation optimized for interference resilience, while the decoder reconstructs the original symbols from the denoised latent vector. This end-to-end optimized system explicitly addresses residual multiuser interference and improves symbol estimation accuracy.

The JSCC encoder \( \mathcal{E}_\phi(\cdot) \) transforms the transmitted QPSK symbol vector \( \mathbf{s}_u \in \mathbb{C}^n \) for UT \( u \) into a latent representation \( \mathbf{z}_u \in \mathbb{R}^{2n} \) (with $n$ being the block size) so that
\begin{equation}\label{eq:jscc_encoder_revised}
\mathbf{z}_u = \mathcal{E}_\phi(\mathbf{s}_u),
\end{equation}
where \( \phi \) represents the encoder's learnable parameters. The encoder employs convolutional neural layers to exploit local correlations in the symbol vector
\begin{equation}\label{eq:encoder_layers_revised}
\mathbf{h}^{(l)} = \sigma\left( \mathbf{W}^{(l)} \mathbf{h}^{(l-1)} + \mathbf{b}^{(l)} \right),~\mbox{for } l = 1, \dots, L,
\end{equation}
with the input \( \mathbf{h}^{(0)} = {\rm Re}(\mathbf{s}_u) \oplus {\rm Im}(\mathbf{s}_u) \), stacking real and imaginary parts, ${\bf W}^{(l)}$ being the weights at the $l$-th layer, ${\bf b}^{(l)}$ being the bias at the $l$-th layer, \( \sigma(\cdot) \) as a leaky ReLU activation and the number of layers, $L$. This architecture ensures that \( \mathbf{z}_u={\bf h}^{(L)} \) captures symbol features while being resilient to channel distortions and interference.

After transmission through the multiuser channel, the received signal after MRC is \( \mathbf{y}_u =[y_u(1),\dots,y_u(n)]^T \), where \( y_u\) is expressed as (\ref{eq:mrc_combine}) where $r_k^{(u)}$ is given in (\ref{eq:rx_signal_sectionII}) but now with $z_u$ replacing $s_u$, which is the encoded latent representation. The MixDDPM denoiser \( \mathcal{D}_\theta(\cdot) \), trained as described in Section~\ref{subsec:mixddpm}, processes \( \mathbf{y}_u \) to form a denoised latent vector
\begin{equation}\label{eq:denoiser_revised}
\hat{\mathbf{z}}_u = \mathcal{D}_\theta(\mathbf{y}_u),
\end{equation}
in which \( \theta \) parameterizes the reverse diffusion process. The denoiser leverages its mixture-based modeling to mitigate the multi-modal interference patterns unique to symbol-level FAMA, refining \( \hat{\mathbf{z}}_u \) for subsequent decoding.

The JSCC decoder \( \mathcal{D}_\psi(\cdot) \) reconstructs the original QPSK symbols from the denoised latent vector, i.e.,
\begin{equation}\label{eq:jscc_decoder_revised}
\hat{\mathbf{s}}_u = \mathcal{D}_\psi(\hat{\mathbf{z}}_u),
\end{equation}
where \( \psi \) represents the decoder's parameters. The decoder mirrors the encoder with transposed convolutional layers
\begin{equation}\label{eq:decoder_layers_revised}
\mathbf{h}^{(l)} = \sigma\left( \mathbf{W}^{(L-l+1)T} \mathbf{h}^{(l-1)} + \mathbf{b}^{(L-l+1)} \right),~\mbox{for }l = 1, \dots, L,
\end{equation}
outputting \( \hat{\mathbf{s}}_u \) in the complex domain by combining the final layer's real and imaginary components.

Overall, the e2e system is trained in order to minimize the mean squared error between the transmitted and reconstructed symbols, augmented by the MixDDPM loss
\begin{align}
\mathcal{L}_{\text{total}} &= \mathbb{E}_{\mathbf{s}_u, \boldsymbol{\eta}_{\text{res}}} \left[ \left\| \mathbf{s}_u - \mathcal{D}_\psi\left( \mathcal{D}_\theta\left( \mathcal{E}_\phi(\mathbf{s}_u) + \boldsymbol{\eta}_{\text{res}} \right) \right) \right\|^2 \right]\\
& \quad\quad\quad\quad\quad\quad\quad\quad+ \lambda \mathcal{L}_{\text{MixDDPM}},\notag\\
&\equiv\mathcal{L}_{\text{JSCC}}+\lambda \mathcal{L}_{\text{MixDDPM}},\label{eq:total_loss_revised}
\end{align}
where \( \lambda \) balances reconstruction accuracy and denoising performance, and \( \mathcal{L}_{\text{MixDDPM}} \) is defined in \eqref{eq:mix_loss}. The parameters \( \phi \), \( \theta \), and \( \psi \) are jointly optimized via gradient descent, enabling the system to adapt to the interference characteristics.

This integration enhances resilience of the proposed FAMA system by combining JSCC's ability to encode interference-robust representations with MixDDPM's capacity to remove structured residual interference. The training procedure for the combined JSCC and MixDDPM is outlined in Algorithm \ref{alg:jscc_mixddpm_combined_corrected}.

\begin{figure}
\begin{algorithm}[H]
\caption{Combined JSCC and MixDDPM Training}\label{alg:jscc_mixddpm_combined_corrected}
\begin{algorithmic}[1]
\Require Training set \( S \), hyperparameters \( T, \alpha_t, \lambda \), CSI at the ports \( \{g^{(u,u)}_k\} \), noise variance \( \sigma_\eta^2 \)
\Ensure Trained JSCC encoder \( \mathcal{E}_\phi \), decoder \( \mathcal{D}_\psi \), and MixDDPM model \( p_\theta \)
\State \textbf{Stage 1: Pre-train JSCC without diffusion}
\While{Training condition not met}
    \State Sample \( \mathbf{s} \sim S \)
    \State Compute latent \( \mathbf{z} = \mathcal{E}_\phi(\mathbf{s}) \)
    \State Sample noise \( \mathbf{n} \sim \mathcal{N}(\mathbf{0}, \sigma_\eta^2 \mathbf{I}) \)
    \State Compute channel output \( \mathbf{y}  \)
    \State Update \( \phi, \psi \) to minimize \( \|\mathbf{s} - \mathcal{D}_\psi(\mathbf{y})\|^2 \)
\EndWhile
\State \textbf{Stage 2: Train MixDDPM}
\While{Training condition not met}
    \State Sample \( \mathbf{x}_0 \sim S \)
    \State Sample timestep \( t \sim \text{Uniform}\{1, \ldots, T\} \)
    \State Sample noise \( \boldsymbol{\epsilon} \sim \mathcal{N}(\mathbf{0}, \mathbf{I}) \)
    \State Compute \( \mathbf{x}_t = \sqrt{\bar{\alpha}_t} \mathbf{x}_0 + \sqrt{1 - \bar{\alpha}_t} \boldsymbol{\epsilon} \)
    \State Sample \( j \sim \text{Categorical}(\omega_1, \ldots, \omega_K) \)
    \State Compute loss:
    \begin{equation*}
        \mathcal{L} = -\log p_\theta(j | \mathbf{x}_t) + \|\boldsymbol{\epsilon} - \boldsymbol{\epsilon}_\theta^{(j)}(\mathbf{x}_t)\|^2
    \end{equation*}
    \State Update \( \theta \) using gradient descent on \( \mathcal{L} \)
\EndWhile
\State \textbf{Stage 3: Joint fine-tuning}
\While{Training condition not met}
    \State Sample \( \mathbf{s} \sim S \)
    \State Compute latent \( \mathbf{z} = \mathcal{E}_\phi(\mathbf{s}) \)
    \State Sample residual interference \( \boldsymbol{\eta}_{\text{res}} \)
    \State Compute received signal $\mathbf{y}$ using (\ref{eq:mrc_combine}) 
    \State Denoise \( \hat{\mathbf{z}} = \mathcal{D}_\theta(\mathbf{y}) \)
    \State Reconstruct \( \hat{\mathbf{s}} = \mathcal{D}_\psi(\hat{\mathbf{z}}) \)
    \State Compute reconstruction loss \( \mathcal{L}_{\text{JSCC}} = \|\mathbf{s} - \hat{\mathbf{s}}\|^2 \)
    \State Compute MixDDPM loss \( \mathcal{L}_{\text{MixDDPM}} \) using \( \theta \)
    \State Update \( \phi, \theta, \psi \) to minimize \( \mathcal{L}_{\text{JSCC}} + \lambda \mathcal{L}_{\text{MixDDPM}} \)
\EndWhile
\end{algorithmic}
\end{algorithm}
\vspace{-8mm}
\end{figure}

\subsection{Model Structure}
The proposed system employs a JSCC encoder that begins by embedding the input data into a latent representation through a series of patch-partition and convolutional layers, followed by multiple blocks integrating local convolutions and transformer-style self-attention to capture both short- and long-range dependencies. The encoded signal $\mathbf{z}$ is then transmitted over the channel, and the receiver applies a diffusion-based denoiser inspired by an improved U-Net architecture \cite{Ronneberger-2015}.

The corrupted signal $\mathbf{y}$ is passed to an initial convolutional layer, which prepares it for the U-Net pipeline \cite{Ronneberger-2015}. Subsequent down-sampling blocks, via stride-based convolutions, reduce spatial resolution while increasing the feature-channel depth. Each U-Net stage consists of a convolutional residual (Conv-Res) block and a convolutional attention (Conv-Attn) block. The Conv-Res block replaces fully-connected layers in the residual path with additional convolutional layers, enhancing feature extraction at reduced computational cost. The Conv-Attn block applies a lightweight attention mechanism using depthwise or grouped convolutions to capture global dependencies. A timestep (or diffusion-step) embedding $t$ is added in the intermediate layers to inform the network about the current stage of the reverse diffusion process.

After the deepest level of the U-Net, up-sampling blocks (interpolation plus a convolutional layer) gradually restore spatial resolution. Skip connections from the encoder stages are concatenated to the decoder stages, enabling the network to recover fine-grained details. The final U-Net output is passed through a projection layer that maps features back into the symbol domain. This denoised signal is then fed to a JSCC decoder, mirroring the encoder's structure but replacing down-sampling modules with the corresponding up-sampling modules to reconstruct the original data. By blending convolution-based feature refinement, attention-driven global modeling, and iterative diffusion, this architecture effectively learns to remove non-Gaussian residual interference characteristics.

\begin{figure*}
\begin{center}
\includegraphics[width=.9\linewidth]{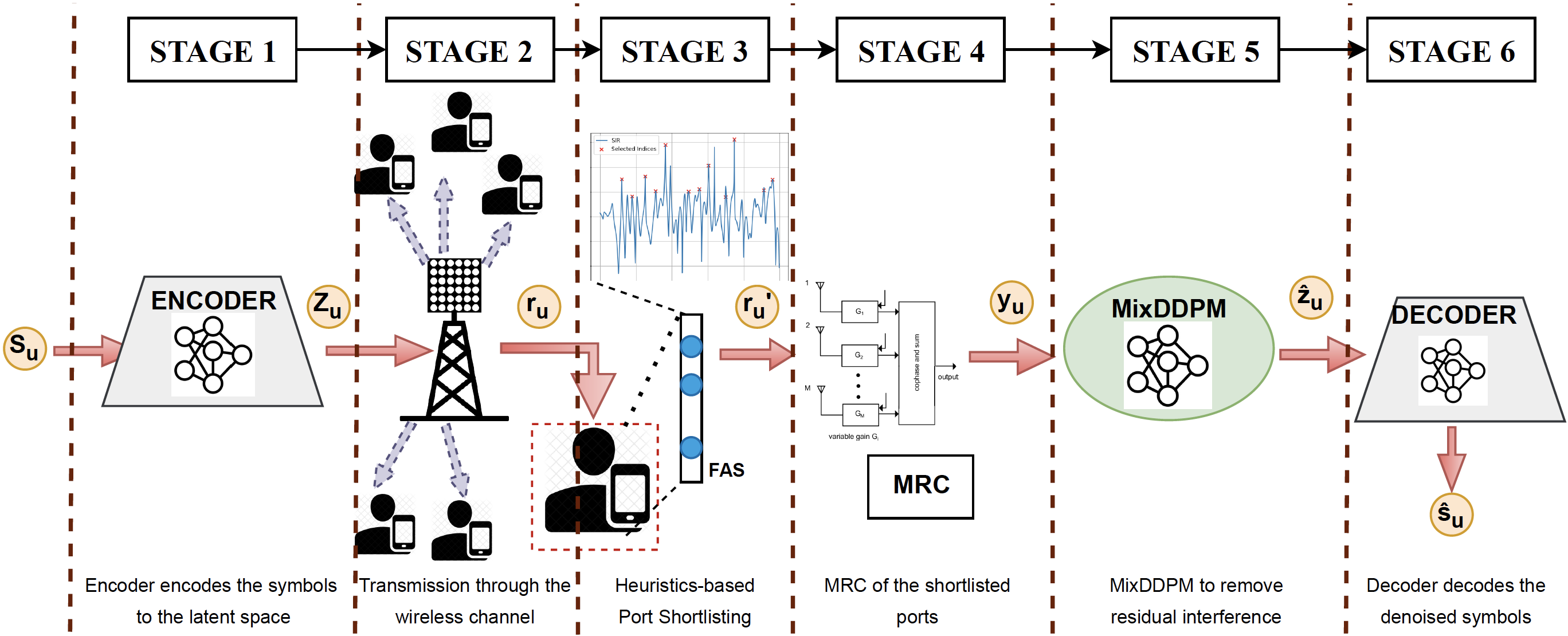}
\caption{The proposed e2e turbo FAMA system with port shortlisting, MRC, MixDDPM and JSCC.}\label{fig:overallsys}
\end{center}
\vspace{-5mm}
\end{figure*}

The overall proposed approach is referred to as turbo FAMA and Fig.~\ref{fig:overallsys} illustrates the blocks of the e2e system.

\vspace{-2mm}
\section{Simulation Results}\label{sec:simulation_results}
In this section, we evaluate the performance of the proposed turbo FAMA system under different settings. We also include two important benchmarks for comparison. All of the methods considered are explained below:
\begin{itemize}
\item Turbo FAMA (with JSCC)---This is the proposed technique that adopts JSCC, port shortlisting, and MRC.
\item Turbo FAMA (with JSCC+MixDDPM)---This is the same as above plus the use of MixDDPM after MRC.
\item All-port MRC---This scheme utilizes all the port received signals for MRC to produce the output signal. Both JSCC and MixDDPM are also used for IUI immunity.
\item Fast FAMA---This scheme assumes perfect knowledge of the port instantaneous SINRs at each UT to find the best port maxmizing it. Also, we consider an excessively large number of ports just for this scheme with $K=1000$. This represents the ideal case for fast FAMA. 
\end{itemize}

Simulation results are conducted under many different settings by varying the number of users, $U$, the number of ports, $K$, and the normalized length of FAS, $W$. Rich scattering Rayleigh fading channels are considered. The average SNR is set to $10~{\rm dB}$ and the blocklength before the encoder is $1024$. We also define the channel bandwidth ratio, ${\rm CBR}\triangleq\frac{n}{1024}$, to measure the bandwidth expansion when utilizing JSCC. If $n=1024$, then no bandwidth expansion is introduced. In the simulations, unless specified otherwise, ${\rm CBR}=1$. Also, in the port shortlisting process, we set $K_{\rm sel}=20$ and $\gamma_\text{th}=0.6$. Note that the parameters for the neural network structures used for JSCC and MixDDPM are listed in TABLE \ref{tab:parameters}.


\begin{table}
\caption{Simulation Parameters}\label{tab:parameters}
\centering
\begin{tabular}{c||c}
\thickhline
{\bf Parameter} & {\bf Value}\\
\thickhline
$T$ (training/sampling) & $1000/50$\\
\hline
mixture weights, $\omega$ & $[0.4,0.15,0.15,0.1,0.1,0.05,0.05]$ \\
\hline
mixture centres, $c$ & $[0, 1, -1, \sqrt{2},-\sqrt{2},\sqrt{3},-\sqrt{3}]$\\
\hline
$\alpha_t$ & $0.9999$ to $0.9800$\\
\hline
Optimizer & ADAM\\
\hline
Learning rate & $0.0001$\\
\hline
dropout & $0.1$\\
\hline
Residual blocks & $2$\\
\hline
Embedding dimensions (enc.) & $[128,256]$\\
\hline
Patch size (enc.) & $2$\\
\hline
Input channels (enc.) & $2$\\
\hline
Attention depth (enc.)& $[6,6]$\\
\hline
Attention heads (enc.)& $[8,16]$\\
\hline
Window size (enc.) & $4$\\
\hline
Embedding dimensions (dec.) & $[256,128]$\\
\hline
Attention depth (dec.)& $[6,6]$\\
\hline
Attention heads (dec.)& $[16,8]$\\
\hline
Window size (dec.) & $4$\\
\hline
Batch size & $64$\\
\hline
Training epochs (JSCC)& $100$\\
\hline
Training epochs (MixDDPM)& $100$\\
\hline
Training epochs (Joint)& $20$\\
\thickhline
\end{tabular}\label{critic}
\vspace{-4mm}
\end{table}

\begin{figure}
\begin{center}
\includegraphics[width=1\columnwidth]{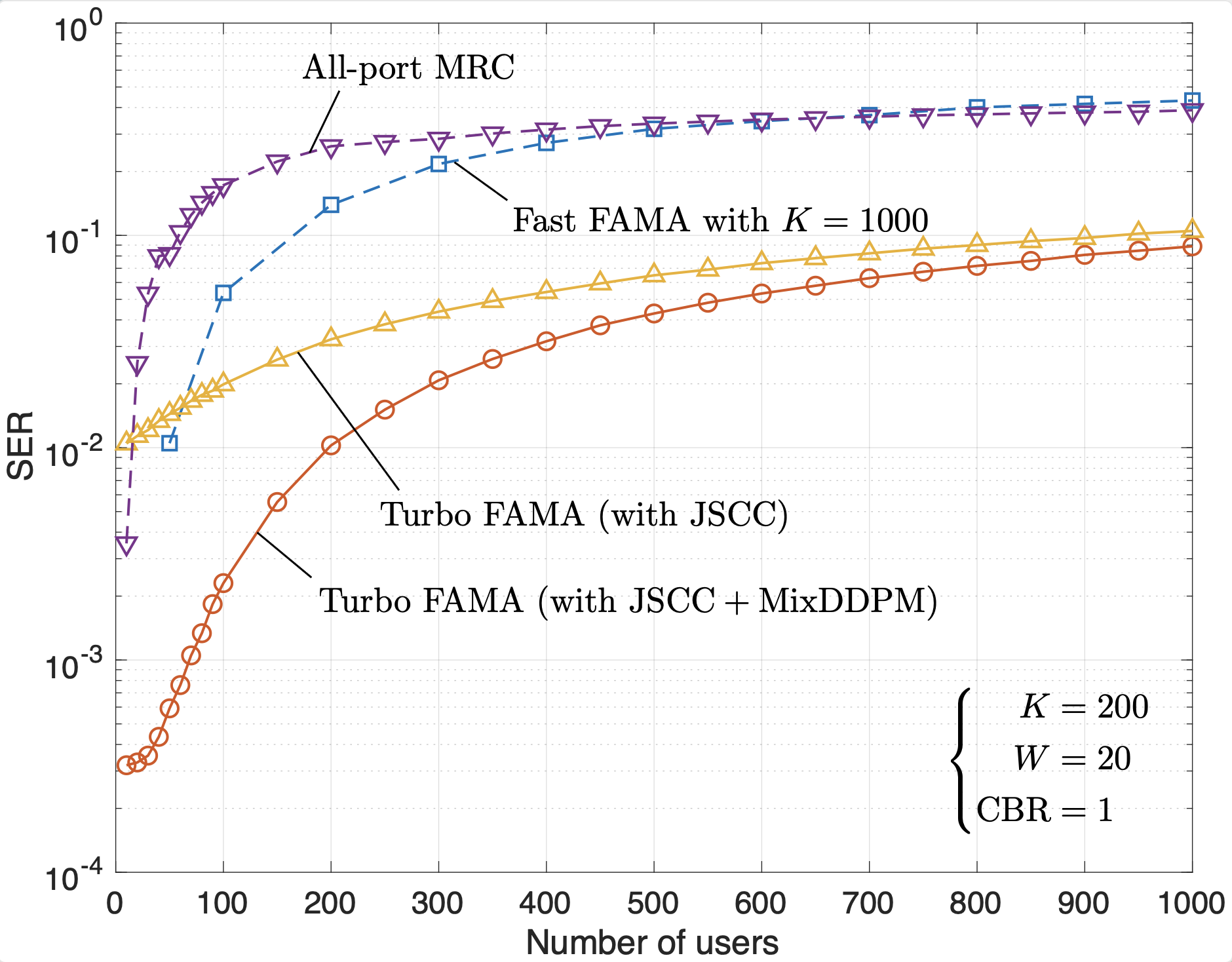} 
\caption{SER versus the number of users for different FAMA approaches with $K=200$, $W=20$, and ${\rm CBR}=1$.}\label{users}
\end{center}
\vspace{-4mm}
\end{figure}

Results in Fig.~\ref{users} are provided for the SER performance for different approaches against the number of users sharing the same physical channel, from a few users to as large as $1000$ users. Note that all the approaches do not require CSI at the BS and no precoding is applied whatsoever. The number of ports at the FAS for each UT is set as $K=200$ except for the fast FAMA method and $W=20$ is assumed. The results illustrate that as the number of users increases, the SER of all the schemes degrades due to elevated IUI, which is expected. Additionally, the SER initially increases sharply, reflecting the challenges in distinguishing desired signals from IUI as the number of users rises in massive scenarios, before gradually plateauing. The results also show that all-port MRC does not really work unless the number of users is small. The ideal fast FAMA scheme does better but is unable to handle more than $200$ users. By contrast, the proposed turbo FAMA approaches outperform both all-port MRC and fast FAMA significantly. The use of MixDDPM in turbo FAMA seems to be particularly effective when the number of users is not so large (say $\le 200$ users) but its effectiveness diminishes if the number of users becomes really large. Remarkably, the results reveal that the proposed turbo FAMA scheme can handle up to $200$ users if ${\rm SER}=0.01$ is required. If a higher SER, say ${\rm SER}=0.1$, is acceptable, then turbo FAMA can even deal with $1000$ users, all without the need of precoding at the BS.


\begin{figure}
\begin{center}
\includegraphics[width=1\columnwidth]{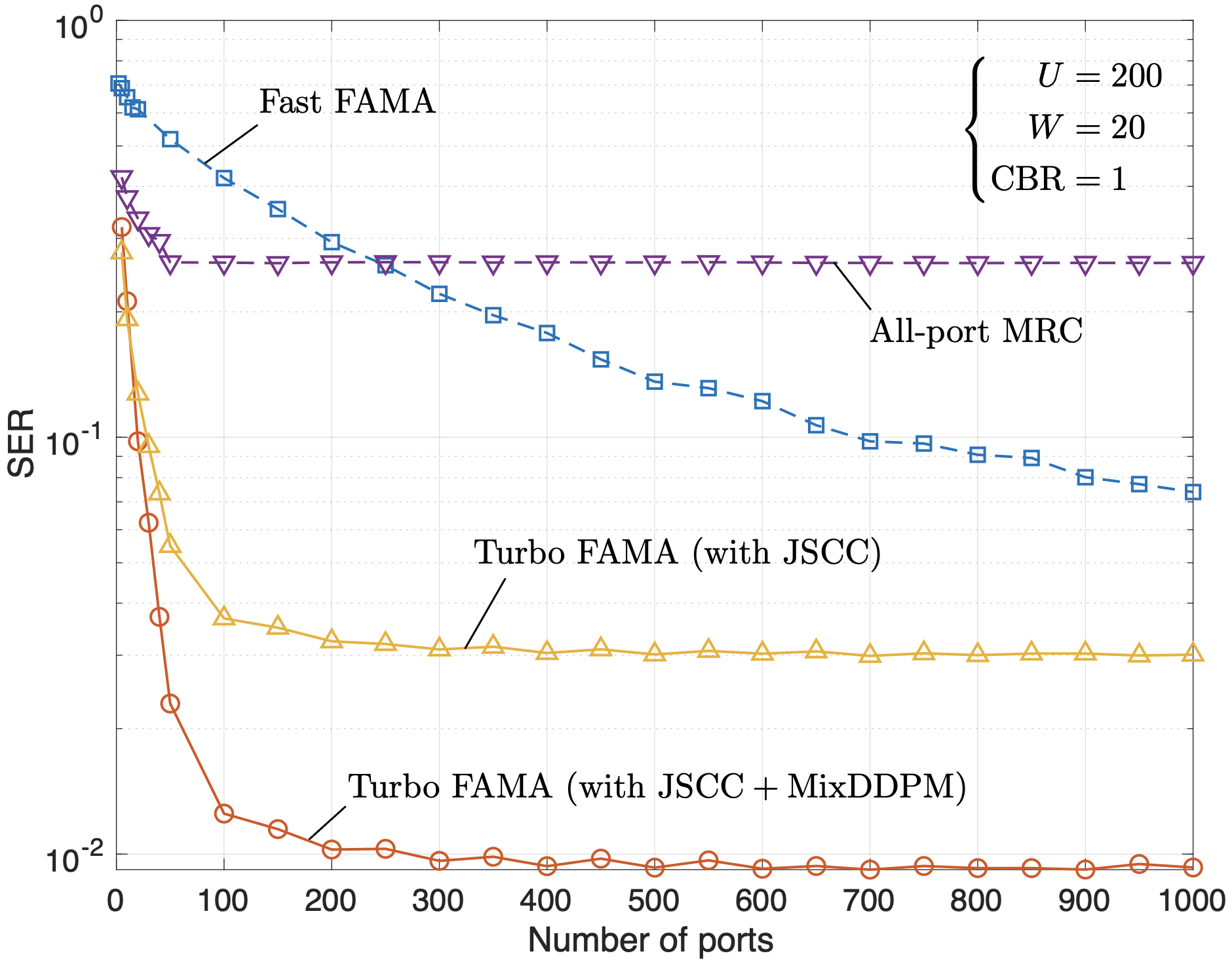} 
\caption{SER versus the number of ports for different FAMA approaches with with $U=200$, $W=20$, and ${\rm CBR}=1$.}\label{ports}
\end{center}
\vspace{-4mm}
\end{figure}

In Fig.~\ref{ports}, we investigate the impact of the number of ports, $K$, on the SER performance for all these FAMA schemes. In the results, we have fixed $U=200$ users while the number of ports is changed. Evidently, we expect that as $K$ increases, the general trend should be that SER decreases. This is clearly what happens to fast FAMA because a larger $K$ will give fast FAMA a higher spatial resolution to avoid IUI. Nevetheless, for the other three methods, all-port MRC and the two turbo FAMA schemes, the SER first drops and then saturates when $K$ reaches a certain point. For all-port MRC, it is clear that higher spatial resolution does not necessarily imply better IUI immunity as MRC does not take into account any knowledge of IUI when combining. For turbo FAMA, on the other hand, it may be explained by the fact that the performance gain of port shortlisting diminishes as $K$ becomes large, meaning that the shortlisted ports do not change much if $K$ continues to increase beyond a certain value. This is in fact a welcoming property because there appears to have an optimal value of $K$ to keep the complexity of training and hardware of FAS low when near-optimal SER performance is achieved. 

\begin{figure}
\begin{center}
\includegraphics[width=1\columnwidth]{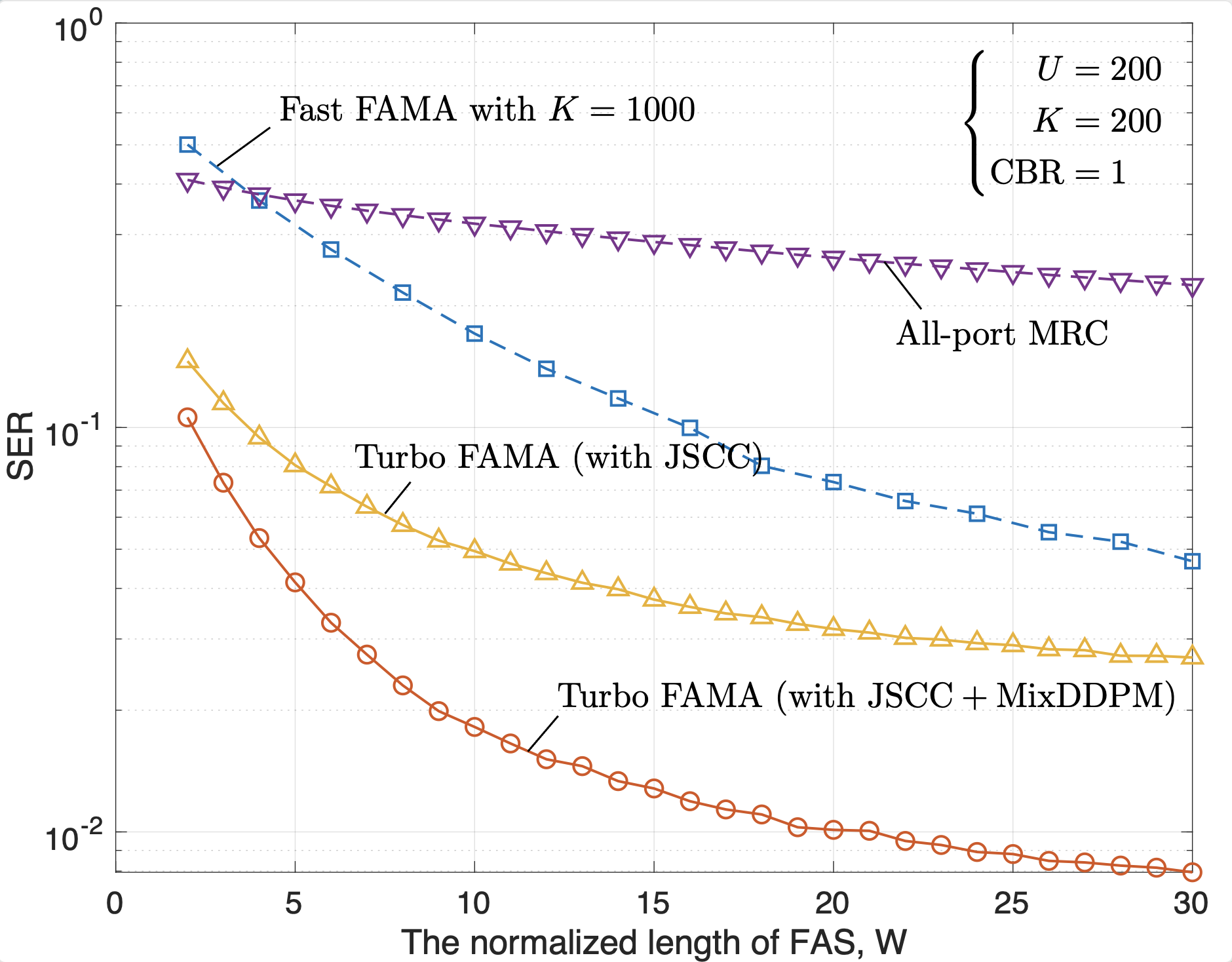} 
\caption{SER versus the normalized size of FAS at each UT for different FAMA approaches with $K=200$, $U=200$, and ${\rm CBR}=1$.}\label{size}
\end{center}
\vspace{-4mm}
\end{figure}

Fig.~\ref{size} illustrates the SER performance versus the normalized size of the fluid antenna at each UT, $W$. As expected, when $W$ increases, the spatial diversity is enhanced because a larger antenna aperture provides more spatial samples, leading to a significant reduction in SER. Moreover, the FAS size affects fast FAMA more than other schemes. In particular, the results suggest that the SER of turbo FAMA saturates more quickly as $W$ increases. This may be explained by the fact that the SDM approach in the port shortlisting process already decorrelates the channels of the shortlisted ports very well even when $W$ is small and thus there is a diminishing return when $W$ becomes large. Additionally, the results demonstrate that MixDDPM is a lot more effective for large $W$. This means that the diffusion-based denoiser relies on the FAS size to work well.


\begin{figure}
\begin{center}
\includegraphics[width=1\columnwidth]{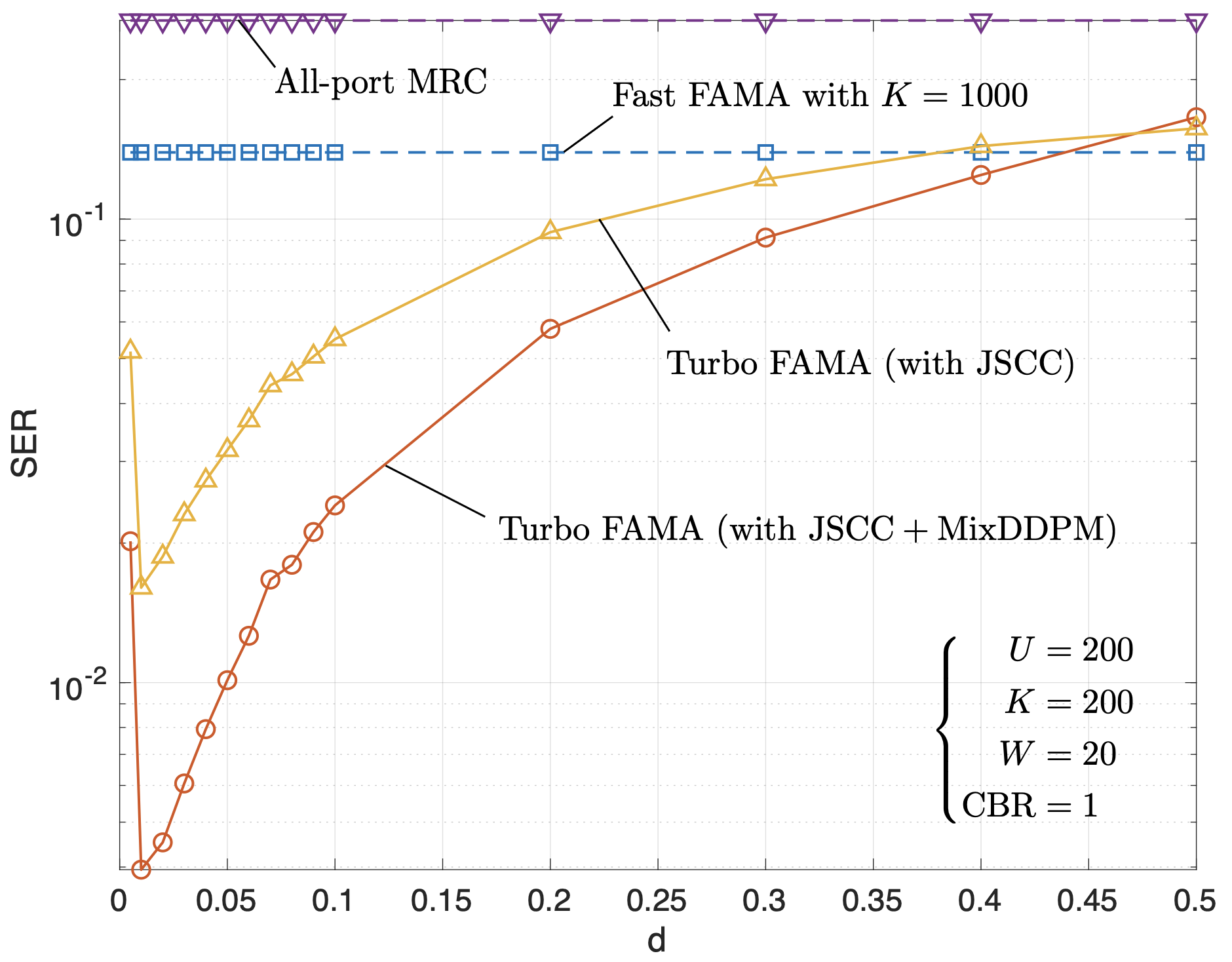} 
\caption{SER versus the spacing between shortlisted ports for different FAMA approaches with $K=200$, $U=200$, $W=20$ and ${\rm CBR}=1$.}\label{distance}
\end{center}
\vspace{-4mm}
\end{figure}

Now, we turn our attention to study the importance of SDM in the port shortlisting process. To do so, in the turbo FAMA schemes, we disable the SDM step but instead enforce a fixed minimum spacing of $d\times K$ between the shortlisted ports from the set ${\cal K}^{(u)}$. The parameter, $d$, allows us to control the spacing between the shortlisted ports and then highlight the importance in the spatial diversity of the shortlisted ports. Note that the spacing parameter, $d$, only affects the proposed turbo FAMA schemes but not all-port MRC and fast FAMA. The corresponding SER results are given in Fig.~\ref{distance}. As we can see, the spacing plays an important role in the SER performance for the turbo FAMA schemes. Specifically, when $d$ increases, the spatial correlation between the shortlisted ports decreases, leading to more independent channel realizations that should facilitate more effective interference mitigation. However, this reduces the possible number of shortlisted ports that degrades the overall performance. This is why for very small values of $d$, although the ports are closely spaced and more correlated, the SER performance is still better. While this may suggest that the smaller the value of $d$, the better the SER performance, it is important to recognize that when $d$ becomes too small, the SER shoots up. This abrupt rise in the SER can be explained by the fact that when $d$ is too small, there are too many closely spaced shortlisted ports with low diversity, and after MRC, gives a degraded signal in terms of IUI. Therefore, there is an optimal value of $d$ which is small but should not be too small for minimizing the SER. For example, the results show that turbo FAMA with MixDDPM (but without SDM) at $d=0.05$ obtains ${\rm SER}=0.01$, the level achieved by turbo FAMA with MxDDPM and SDM. If $d=0.01$, then turbo FAMA achieves the minimum SER of $0.004$. However, finding the optimal $d$ is tedious and requires retraining of JSCC and MxDDPM. The use of SDM thus can be viewed as a smart compromise, which balances between diversity and the number of shortlisted ports, and eliminates the need of choosing $d$.


\begin{figure}
\begin{center}
\includegraphics[width=1\columnwidth]{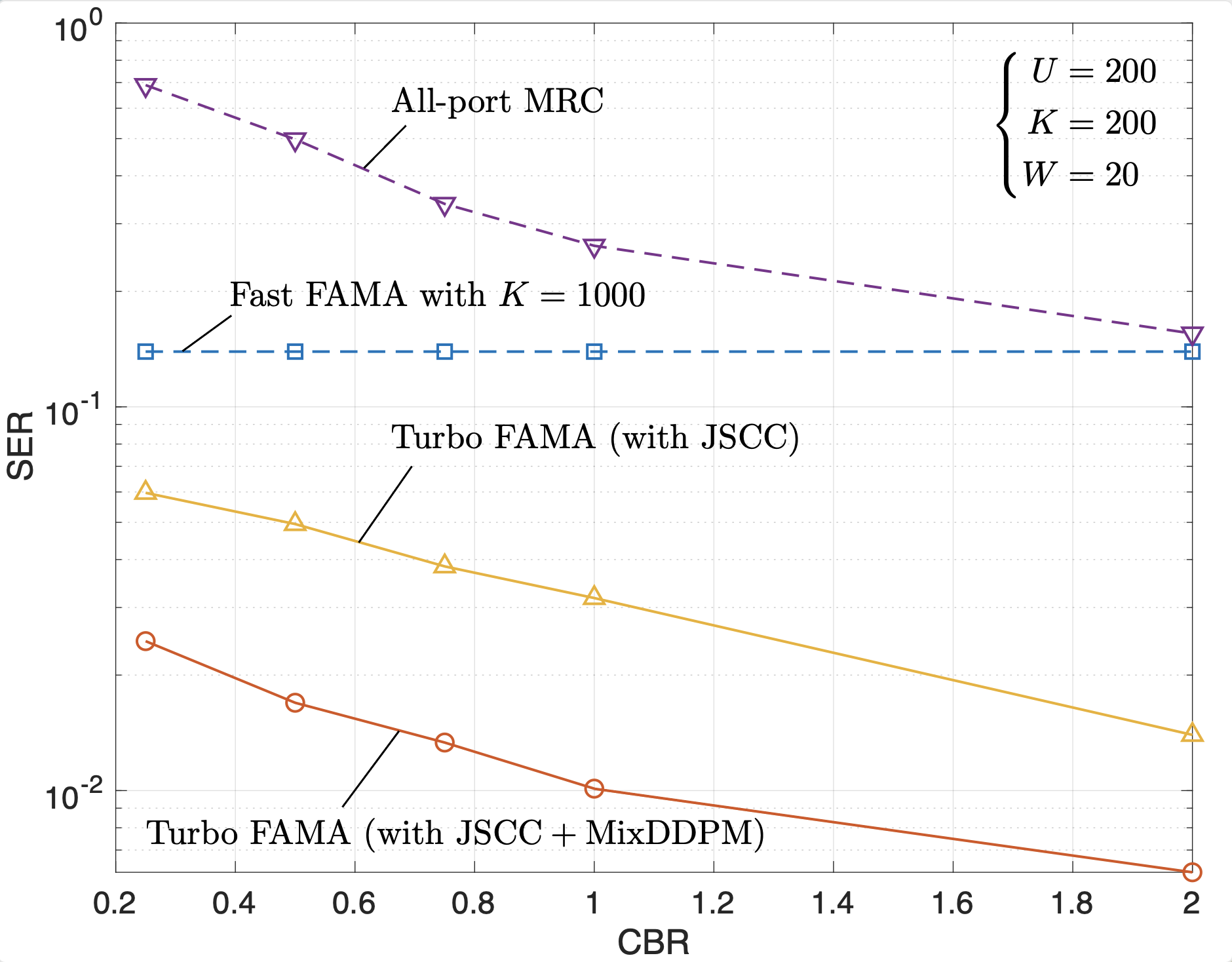} 
\caption{SER versus the CBR for different FAMA approaches with $K=200$, $U=200$, and $W=20$.}\label{CBR}
\end{center}
\vspace{-4mm}
\end{figure}

When using JSCC, it is actually possible to either introduce redundancy for improved error correction or compression for bandwidth saving. This effect can be considered by considering different values of CBR. Note that JSCC is employed for all-port MRC and the two turbo FAMA schemes except fast FAMA. The results in Fig.~\ref{CBR} illustrate that the SER decreases as the CBR increases, which will make all-port MRC approach the performance of fast FAMA. Additionally, it is possible to achieve bandwidth saving using JSCC in the proposed turbo FAMA schemes with some mild SER degradation.

\begin{figure}
\begin{center}
\includegraphics[width=1\columnwidth]{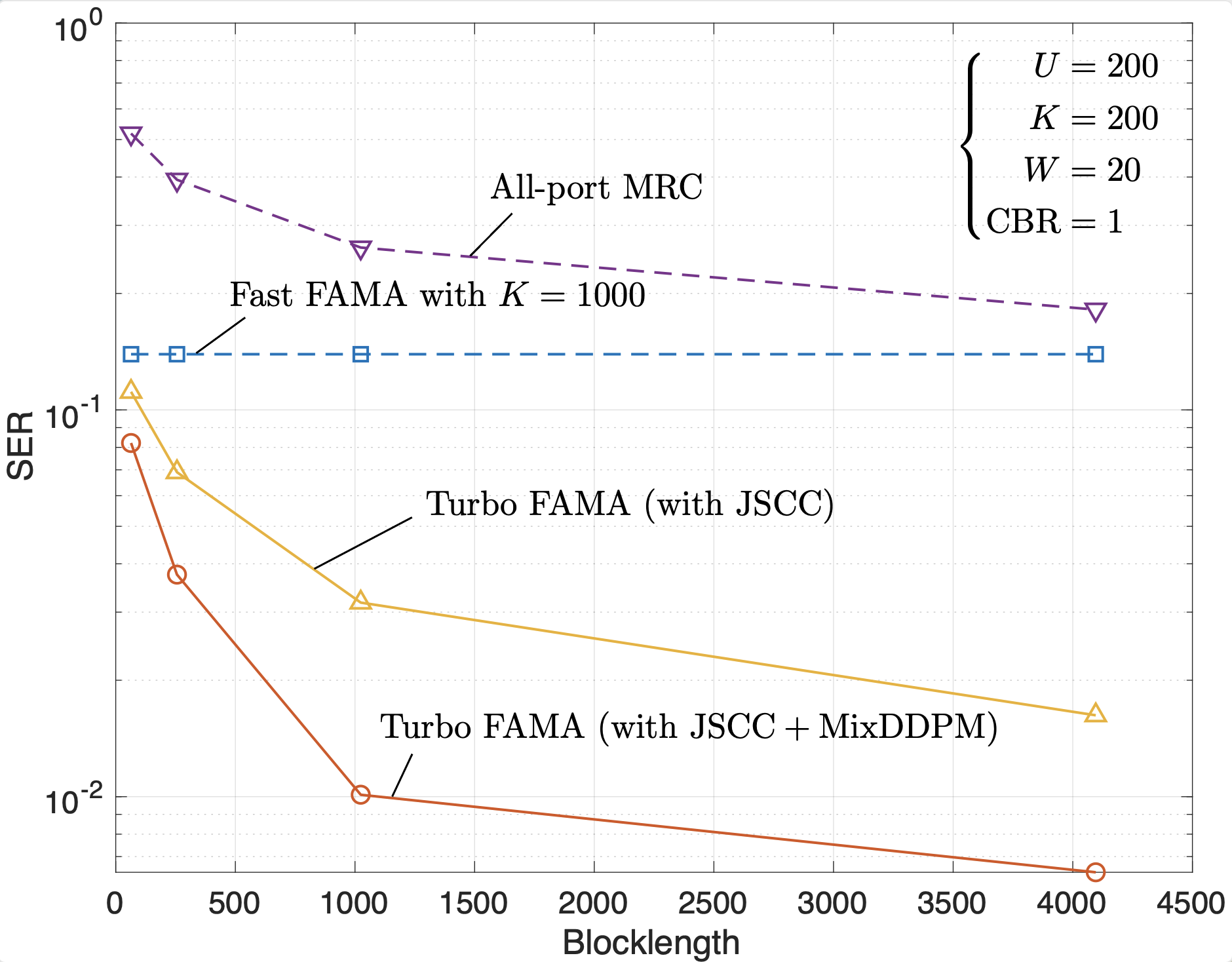} 
\caption{SER versus the blocklength for different FAMA approaches with $K=200$, $U=200$, $W=20$ and ${\rm CBR}=1$.}\label{blocklength}
\end{center}
\vspace{-4mm}
\end{figure}

Finally, we provide results to investigate how SER changes according to the blocklength of the data inputting to the JSCC encoder. As before, these results only affect all-port MRC and the two turbo FAMA schemes as fast FAMA does not apply JSCC. These results are provided in Fig.~\ref{blocklength}, showing that as the blocklength increases, all schemes exhibit improved SER. However, beyond a certain point, the SER improvement tends to saturate. This saturation can be attributed to diminishing returns in redundancy gains without increasing CBR, as well as limitations in training complexity. 

\vspace{-2mm}
\section{Conclusion}\label{sec:conclusion}
In this paper, we proposed a new FAMA scheme that utilizes position flexibility in FAS at UTs to realize extreme massive access, without the need for precoding at the BS. The proposed scheme is dubbed turbo FAMA as it integrates heuristic port shortlisting, MRC, JSCC and diffusion-based denoising using MixDDPM into one. The overall turbo FAMA system is able to shortlist promising ports for obtaining favourable received signals for MRC, which is followed by MixDDPM denoising to suppress non-Gaussian residual interference, and then JSCC decoding. Our simulation results point to a promising future, showing that it is possible to support $200$ users on the same channel using turbo FAMA if ${\rm SER}=0.01$ is required. Further, if only ${\rm SER}=0.1$ is needed, the proposed turbo FAMA can even serve $1000$ users, truly answering the scalability issue in massive access scenarios. Future work should look at ways to reduce the complexity of obtaining the received signals at all the FAS ports, and study the performance impact if only some of the received signals at the FAS are available.

\ifCLASSOPTIONcaptionsoff
  \newpage
\fi

\bibliographystyle{IEEEtran}

\end{document}